\documentclass[12pt] {article}

\usepackage{amsmath}
\usepackage{epsfig}
\usepackage{setspace}
\usepackage{cite}
\usepackage{color}
\usepackage{subfigure}

\begin{document}

\title{Mesoscopic effective material parameters for thin layers modeled as single and double grids of interacting
loaded wires}

\author{Pekka Ikonen$^{1}$, Elena Saenz$^{2,1}$, Ramon Gonzalo$^{2}$,\\Constantin Simovski$^{1}$, and Sergei Tretyakov$^{1}$}

\date{$^{1}$Radio Laboratory/SMARAD Centre of Excellence\\TKK Helsinki University of
Technology\\P.O. Box 3000, FI-02015 TKK, Finland\\ \vspace{0.5cm} $^2$Department of Electrical and Electronic
Engineering\\Public University of Navarra\\Campus Arrosadia E-31006, Pamplona-Navarra, Spain}

\maketitle

{\center \large

Address for correspondence:

Sergei Tretyakov\\ Radio Laboratory, TKK\\ P.O.~Box 3000, FI-02015 TKK, Finland

Fax: +358-9-451-2152

E-mail: sergei.tretyakov@tkk.fi

}

\maketitle


\begin{center}
\section*{Abstract}
\end{center}

As an example of thin composite layers we consider single and double grids of periodically arranged interacting
wires loaded with a certain distributed reactive impedance. Currents induced to the wires by a normally incident
plane wave are rigorously calculated and the corresponding dipole moment densities are determined. Using this
data and the averaged fields we assign mesoscopic material parameters for the proposed grid structures. These
parameters depend on the number of grids, and measure the averaged induced polarizations. It is demonstrated
that properly loaded double grids possess polarization response that over some frequency range can be described
by assigning negative values for the mesoscopic parameters. Discussion is conducted on the physical
meaningfulness to assign such material parameters for thin composite slabs. The results predicted by the
proposed method for the double-grid structures are compared with the results obtained using the commonly adopted
$S$-parameter retrieval procedure.

\newpage

\section{Introduction}

The present work is motivated by recent articles in which the authors assign very exotic material parameters
$\epsilon$, $\mu$ (or effective refractive index $n$) for thin composite slabs consisting only of a few grids of
interacting complex-shaped inclusions, e.g.~\cite{Linden,Dolling,Shalaev,Zhou}. In all of the above referred
works the effective material parameters have been extracted using the $S$-parameter retrieval procedure whose
underlying basics are described e.g.~in \cite{Smith1, Chen, Smith2}. In this procedure one replaces the actual
structure by a slab of homogeneous material having a certain effective thickness and searches for effective
material parameters that will reproduce the scattering results obtained with a normal plane-wave incidence. The
method is often known to lead to antiresonant behavior for one of the extracted material parameters, which
further leads to physically wrong sign for the imaginary part of the corresponding material parameter
\cite{Koschny}. According to the classical electrodynamics \cite{Landau}, a physically meaningful material
parameter is a linear response that must be independent of the electromagnetic field phasors excited in the
medium, and especially in case of composite media, independent of the geometrical construction constituting the
medium. Moreover, the causality requirement (which is directly related to the fulfillment of the Kramers-Kroning
relations) and the passivity requirement (which relates to the allowed sign of the imaginary part) must be
simultaneously fulfilled \cite{Landau, Depine, Efros, Costia}. It is therefore clear that an effective parameter
showing antiresonant behavior does not bear the meaning of a physically sound material parameter describing a
real homogeneous material\footnote{Please note that the function typically fitting into the observed
antiresonant frequency dependence \cite{Koschny} resembles the Lorentz model with the sign in front of the
fractional term changed (we assume $e^{j\omega t}$ time dependence):$$F(\omega) = 1 -
\frac{A\omega^2}{\omega_0^2 - \omega^2 + j\omega \Gamma}.$$ The change of the sign assures that the poles of the
function lie in the upper half-omega plane (thus the function can be said to be causal), however, the sign of
the real part of frequency derivative is wrong to fulfill the passivity requirement.}, it only allows to
reproduce the scattering scenario with a specific excitation \cite{Koschny2}. When adopting the aforementioned
classical definition for physically sound material parameters it is clear that, according to this definition,
the parameters presented e.g.~in \cite{Linden, Dolling, Zhou} are nonphysical. Also, in the view of classical
homogenization approaches (e.g.~\cite{Sihvola}) it is clear that a structure consisting only of a few (e.g.~two)
interacting grids of inclusions does not correspond to any slab of an effectively homogeneous material. In an
effectively homogeneous material the unit cell should contain a large number of molecules or inclusions, and yet
still remain small compared to the wavelength. This partly assures e.g.~that the change of the material sample
thickness does not alter the polarization response of a small material volume. Evidently, the material
parameters assigned for thin composite slabs are \emph{mesoscopic}, thus, they depend on the number of
monolayers (grids) constituting the sample. Most metamaterials and thin composite layers operate in the
mesoscopic regime, intermediate between the homogeneous medium regime, and the regime of a collection of weakly
interacting individual scatterers.

Typically the authors using the $S$-parameter retrieval procedure explain that the finite spatial periodicity of
the sample causes the often observed antiresonant behavior. It is also well known that the method introduces an
ambiguity in the phase of the wave reflected from finite thickness structures. To avoid this ambiguity, attempts
have been made to assign effective material parameters for finite thickness complex slabs through refraction,
rather than through reflection. For example, in \cite{Schwartz} the authors a present clear discussion on the
challenges to assign physically sound effective refractive index for wire medium slabs having a finite
thickness. A small artificial interface shift is introduced for the actual structure and the authors use an
iterative minimization technique to find the effective refractive index of a homogeneous slab that transmits the
same amount of light as the actual complex structure (essentially the same technique was used also e.g.~in
\cite{Guida}). Authors of \cite{Schwartz} also define the refractive index through reflection and transmission,
but they search for the material parameters that allow one to reproduce the actual scattering results with all
incidence angles (note that the technique introduced e.g.~in \cite{Smith1} uses only reflection and transmission
data for a normally incident plane wave).

Motivated by the recent contradictory dispersion results obtained for single and double grids of interacting
inclusions \cite{Linden, Dolling, Zhou} we turn our attention to single and double grids of loaded wires. Grids
of loaded wires are considered since the theory of interacting grids of such wires is well established and
accurate \cite{Pavel, Pavel_PRE}, and certain specific loading scenarios allow us to simulate the dispersion
properties of lattices of small dipole scatterers (more discussion on this will be presented later). Also,
despite the fact that the local field approach used in this paper completely takes into account the near-field
interactions (all the higher-order Floquet harmonics) between the wires, is remains physically illustrative and
computationally efficient. Our goal is to assign mesoscopic material parameters for the proposed grids of wires
directly by solving the induced dipole moment densities and averaged fields. Recently we used the same approach
for assigning effective material parameters for grids of interacting inclusions modeled as small electric dipole
scatterers \cite{Elena}. In this work our goal is, in addition to studying different structures as considered in
\cite{Elena}, present a more extensive discussion on the physics behind the obtained polarization responses, and
more systematically compare (with double grid structures) the proposed results to the results obtained using the
$S$-parameter retrieval procedure. The details of derivation are introduced, and the obtained results and their
physical meaning are thoroughly discussed.

\section{Formulation}

\subsection{Outline of the strategy}

The known constitutive relations in a certain macroscopic medium (or in a large sample of a certain macroscopic
material) read \cite{Jackson}
\begin{equation}
\widehat{\textbf{D}} = \epsilon_0\widehat{\textbf{E}} + \textbf{P}, \label{D}
\end{equation}
\begin{equation}
\widehat{\textbf{B}} = \mu_0\widehat{\textbf{H}} + \textbf{M}, \label{B}
\end{equation}
where $\widehat{\textbf{D}}$ and $\widehat{\textbf{B}}$ are the averaged electric and magnetic flux density,
$\widehat{\textbf{E}}$ and $\widehat{\textbf{H}}$ are the averaged electric and magnetic field, $\textbf{P}$ is
the averaged polarization (electric dipole moment density), and $\textbf{M}$ is the averaged magnetization
(magnetic dipole moment density). When introducing a physically small averaging volume $V$, the average
polarization and magnetization can be calculated as \cite{Jackson}
\begin{equation}
\textbf{P} = \frac{\textbf{p}}{V} = \frac{1}{j\omega  V}\int \textbf{i}(r') \hspace{0.15cm} d^3r', \label{P}
\end{equation}
\begin{equation}
\textbf{M} = \frac{\textbf{m}}{V} = \frac{\mu_0}{V}\int \textbf{r}'\times \textbf{i}(r') \hspace{0.15cm} d^3r',
\label{M}
\end{equation}
where $\textbf{i}$ is the microscopic current density confined in the averaging volume, and the integration is
done over that volume.

In the following analysis our strategy is to first solve the current induced to the wires by the incident field,
taking into account the full interaction between all the wires. After this, we define an artificial unit volume
and calculate the macroscopic polarization and magnetization using eqs.~(\ref{P}) and (\ref{M}). Further, the
total averaged fields in the unit volume are defined, and eqs.~(\ref{D}) and (\ref{B}) are used to express the
mesoscopic material parameters in terms of macroscopic polarization (magnetization) and averaged fields.

\subsection{Single grid of loaded wires}

\begin{figure}[b!]
\centering \epsfig{file=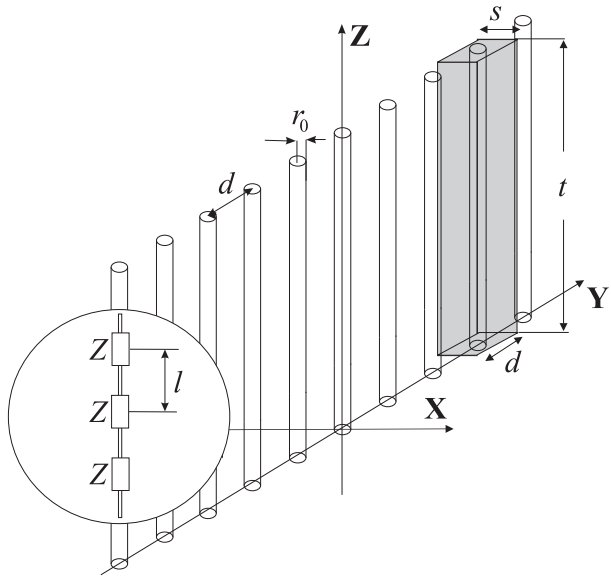, width=8cm} \caption{\emph{Grid of thin wires loaded with bulk impedances
($Z=0$ corresponds to a grid on unloaded wires). The wires are infinitely long and the number of wires in the
$y$-direction is infinite. The insertion period for the loads $l$ is much smaller than the wavelength. An
example unit cell is shown shadowed.}} \label{grid}
\end{figure}

Consider a periodical grid of identically loaded, infinitely long wires depicted schematically in
Fig.~\ref{grid}. The radius of wires $r_0$ is assumed to be small enough so that the wires can be represented as
infinitely long current lines located at the position of the wires axis. Throughout the rest of the paper we
consider normal plane-wave incidence and express the external electric field as
\begin{equation}
\textbf{E}^{\rm ext} = \textbf{u}_zE^{\rm ext}e^{-jkx}. \label{Eext}
\end{equation}

Let us first assume the wires to be unloaded and pick up a reference wire from the grid. The relation between
the amplitude of the local electric field $E^{\rm loc}_z$ acting on the surface of the reference wire and
current $I$ flowing on the wire surface reads
\begin{equation}
E^{\rm loc}_z = \alpha_0^{-1}I, \label{Eloc1}
\end{equation}
where $\alpha_0$ is the susceptibility of the wire which can be determined from the boundary condition on the
wire surface (see e.g.~\cite{Felsen}):
\begin{equation}
\alpha_0 = \bigg{[}\frac{\eta k}{4}H_0^{(2)}(kr_0)\bigg{]}^{-1}, \label{alpha_0}
\end{equation}
where $\eta$ is the wave impedance in the background medium, and $H_0^{(2)}$ represents the Hankel function of
the second kind and zero order. If the wires are loaded with a certain uniformly distributed impedance per unit
length $Z$, the inverse of the wire susceptibility becomes \cite{Pavel_PRE, Sergei}:
\begin{equation}
\alpha^{-1} = \alpha_0^{-1} + Z.
\end{equation}
In the following we denote the wire susceptibility generally as $\alpha$ ($\alpha_0$ is a special case for the
susceptibility when $Z=0$). The amplitude of the local electric field acting on the reference wire can also be
expressed in the following manner:
\begin{equation}
E^{\rm loc} = E^{\rm ext} + \beta(0)I, \label{Eloc2}
\end{equation}
where $E^{\rm ext}$ is the amplitude of the external field at the surface of the reference wire and $\beta(0)$
is so called self interaction coefficient given by eq.~(11) in \cite{Pavel} (see also \cite{Yatsenko}):
\begin{equation}
\beta(0) = -\frac{\eta k}{2}\bigg{[}\frac{1}{kd} - \frac{1}{2} +\frac{j}{\pi}\bigg{(}\log\frac{kd}{4\pi} +
\gamma\bigg{)} +\frac{j}{d}\sum_{n\neq0}\bigg{(}\frac{1}{\sqrt{(2\pi n/d)^2 - k^2}} - \frac{d}{2\pi
|n|}\bigg{)}\bigg{]}, \label{b0}
\end{equation}
where $d$ is the separation of the wires in the grid, and $\gamma\approx0.5772$ is the Euler constant.
Physically $\beta(0)$ tells how strongly the wires located in the same grid as the reference wire contribute to
the amplitude of the local field. From eqs.~(\ref{Eloc1}) and (\ref{Eloc2}) we can solve the amplitude of
current induced to the reference wire as
\begin{equation}
I = \frac{E^{\rm ext}}{\alpha^{-1} - \beta(0)}. \label{I}
\end{equation}

Next, let us turn our attention to the definition of the material parameters. Let the unit cell volume ($x\times
y\times z$) shown shadowed in Fig.~\ref{grid} be $V=s\times d\times l$. 
When the integration in eq.~(\ref{P}) is conducted, the average polarization reads:
\begin{equation}
\textbf{P} = \textbf{u}_z\frac{I}{j\omega sd} = \frac{\widehat{\textbf{J}}}{j\omega s}, \label{intf}
\end{equation}
where we have defined the unit cell averaged surface current density $\widehat{\textbf{J}}=\textbf{u}_zI/d$. We
proceed on defining the total averaged electric\footnote{In the case of a single grid of wires we do not need to
consider the averaged magnetic field, as due to the thin wire assumption there is no circulating current, and
the amplitude of the induced magnetic dipole moment density is zero.} field in the grid plane. In the far zone,
the electric field scattered by the grid (in the normal direction) is a plane-wave field, and close to the grids
the amplitude of this field can be expressed as \cite{Sergei}
\begin{equation}
\textbf{E}^{\rm sc}=-\frac{\eta}{2}\widehat{\textbf{J}}. \label{Esc}
\end{equation}
The far field reflection and transmission coefficients (referred to the grid plane) can be formulated using the
ratio between the scattered and external electric fields \cite{Sergei}
\begin{equation}
R = \frac{E^{\rm sc}}{E^{\rm ext}}, \quad T = 1 + \frac{E^{\rm sc}}{E^{\rm ext}}. \label{RT}
\end{equation}
By definition [22,~p.~79], the total averaged electric field in the grid plane is the sum of the external field
and the plane wave field created by the grid:
\begin{equation}
\textbf{E} = \textbf{E}^{\rm ext} - \frac{\eta}{2}\widehat{\textbf{J}}. \label{Etot}
\end{equation}
To calculate the averaged field we need to average $\textbf{E}$ over the unit cell thickness $s$ (since we
consider plane waves the only variation in the fields occurs in $x$-direction). The final result reads:
\begin{equation}
\widehat{\textbf{E}} = \frac{\sin(ks/2)}{ks/2}\textbf{E}^{\rm ext} - \frac{1 -
e^{-jks/2}}{jks/2}\frac{\eta}{2}\widehat{\textbf{J}}. \label{Eav}
\end{equation}
When the unit cell thickness $s$ tends to zero the limit obviously reduces to eq.~(\ref{Etot}). By inserting
eqs.~(\ref{intf}) and (\ref{Eav}) into (\ref{D}) and dropping out the vector notation we get the following
expression for the mesoscopic effective permittivity:
\begin{equation}
\epsilon = 1 + \frac{k\widehat{J}}{j\omega\epsilon_0[2\sin(ks/2)E^{\rm ext} + j\eta\widehat{J}(1 -
e^{-jks/2})]}. \label{epseff1_m}
\end{equation}
When we use relation (\ref{I}) to express $\widehat{J}$ as a function of $E^{\rm ext}$ another expression is
obtained:
\begin{equation}
\epsilon = 1 + \frac{1}{2d/\eta[\alpha^{-1} - \beta(0)]\sin(ks/2) + e^{-jks/2} - 1}. \label{epseff1}
\end{equation}



One should bear in mind that the current induced to the grid varies as a function of the incidence angle, thus,
different polarization response is expected to occur when the incidence angle varies \cite{Yatsenko, Sergei}.
Importantly we also observe that for the \emph{oblique incidence} also higher order Floquet harmonics contribute
to the scattered electric field calculated over the unit volume thickness. Neglecting the effect of these
harmonics would leave us with non-local mesoscopic permittivity \cite{Costia}.

The fact that the above deduced permittivity is not independent of the field phasors excited in the slab
(incidence angle), or the geometrical construction (thickness) constituting the slab clearly indicates that the
permittivity function does not fulfill the classical criteria set for physically sound material parameters of a
real homogeneous material \cite{Landau}. Nevertheless, when accepting this fact and bearing in mind the
mesoscopic nature of the deduced parameter, the permittivity expression is still useful in studying practically
engineered slabs (although they are not sufficient for enough complete description of the composite):~correctly
deduced mesoscopic material parameters contain physically sound information about the actual (microscopic)
polarization response of the structure.

\subsection{Double grid of loaded wires}

In this subsection we consider two grids of wires as shown in Fig.~\ref{2grid}. In this case at some frequency
(and with proper impedance loading) currents in the grids will be out of phase giving rise to a rather strong
circulating current and induced magnetic dipole moment density (this will be demonstrated using a specific
example later). Following the reasoning in the previous subsection we first write down the system of equations
for the unknown induced currents:
\begin{figure}[b!]
\centering \epsfig{file=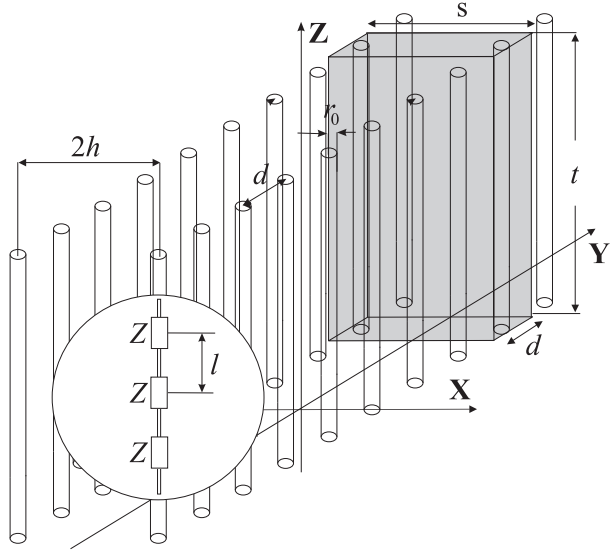, width=8cm} \caption{\emph{Two grids of wires loaded with bulk impedances
($Z=0$ corresponds to grids of unloaded wires). The wires are infinitely long and the number of wires in the
$y$-direction is infinite. The insertion period for the loads $l$ is much smaller than the wavelength. An
example unit cell is shown shadowed.}} \label{2grid}
\end{figure}
$$
\alpha^{-1}I_1= E^{\rm ext}e^{jkh} + \beta(0)I_1 + \beta(2h)I_2,
$$
\begin{equation}
\alpha^{-1}I_2= E^{\rm ext}e^{-jkh} + \beta(2h)I_1 + \beta(0)I_2, \label{virrat}
\end{equation}
where $\beta(2h)$ is so-called mutual interaction coefficient (eq.~(12) in \cite{Pavel}) taking physically into
account the mutual interaction between the grids:
\begin{equation}
\beta(2h) = -\frac{\eta(k^2 - k_z^2)}{2kd}\sum_{n=-\infty}^{n=+\infty}\frac{e^{-jk_x^{(n)}2h}}{k_x^{(n)}}.
\end{equation}
Above $k_x^{(n)}$ is the $z$-component of the $n$-th Floquet mode wave vector \cite{Pavel}:
\begin{equation}
k_x^{(n)} = -j\sqrt{(k_y + 2\pi n/d)^2 + k_z^2 - k^2}, \quad \rm{Re}\{\sqrt{()}\}>0.
\end{equation}

Currents in the grids can be found by solving the system of equations (\ref{virrat}). The average polarization
and magnetization take the following forms after applying eqs.~(\ref{P}) and (\ref{M}) (we use the notation
introduced in the previous subchapter for the unit cell dimensions):
\begin{equation}
\textbf{P} = \frac{\widehat{\textbf{J}}_1 + \widehat{\textbf{J}}_2}{j\omega s},
\end{equation}
\begin{equation}
\textbf{M} = \mu_0\frac{h}{s}({\widehat{\textbf{J}}_2 - \widehat{\textbf{J}}_1}).
\end{equation}

Both grids will create a scattered electric field whose amplitude is given by eq.~(\ref{Esc}). When the electric
field is averaged over the unit cell thickness the result reads:
\begin{equation}
\widehat{\textbf{E}} = \bigg{[}\textbf{E}^{\rm ext}\sin(ks/2) + j\frac{\eta}{2}(\widehat{\textbf{J}}_1 +
\widehat{\textbf{J}}_2)(1 - \cos(kh)e^{-jks/2})\bigg{]}\frac{1}{ks/2}. \label{E2ave}
\end{equation}
The expression for the averaged magnetic field reads:
\begin{equation}
\widehat{\textbf{H}} = \frac{1}{\eta}\bigg{[}\textbf{E}^{\rm ext}\sin(ks/2) -
\frac{\eta}{2}(\widehat{\textbf{J}}_1 - \widehat{\textbf{J}}_2)\sin(kh)e^{-jks/2}\bigg{]}\frac{1}{ks/2}.
\label{H2ave}
\end{equation}
Further, the mesoscopic effective material parameters are determined by applying eqs.~(\ref{D}) and (\ref{B}).
The final expressions read (in the double-grid case the expressions become quite cumbersome when the current
densities are substituted as functions of the external field. For this reason only expressions containing the
current densities are presented):
\begin{equation}
\epsilon = 1 + \frac{k(\widehat{J}_1 + \widehat{J}_2)}{j\omega\epsilon_0\{2E^{\rm ext}\sin(ks/2) +
j\eta(\widehat{J}_1 + \widehat{J}_2)[1 - \cos(kh)e^{-jks/2}]\}},
\end{equation}
\begin{equation}
\mu = 1 + \frac{kh\eta(\widehat{J}_2 - \widehat{J}_1)}{2E^{\rm ext}\sin(ks/2) - \eta(\widehat{J}_1 -
\widehat{J}_2)\sin(kh)e^{-jks/2}}.
\end{equation}

\section{Example results}

Let us choose the following structural parameters for the wire grids (see Figs.~\ref{grid} and \ref{2grid}):
$r_0=0.1$ mm, $d=200r_0$. The insertion period for the loads $l=50r_0$. In the double-grid case the separation
between the grids $2h=40r_0$. These parameters are kept the same throughout the rest of the paper.

We consider three different loading scenarios:
\begin{enumerate}
\item the wires in the grid(s) are unloaded ($Z=0$),
\item the wires are loaded with bulk capacitors [$Z=1/(j\omega
Cl)$],
\item the wires are loaded with a parallel connection of bulk capacitors and inductors [$1/Z = l(j\omega
C + 1/(j\omega L))^{-1} $].
\end{enumerate}
Please note that we do not separately consider the loading of the wires with bulk inductors or with a series
connection of bulk capacitors and inductors. These two situations effectively resemble the scenarios described
in items 1) and 2), as the effect of inductive loading in series can be attributed to the change of the
effective wire radius \cite{Pavel_PRE}.

\subsection{Single grid of wires}

\subsubsection{Unloaded wires}

\begin{figure}[b!]
\centering \epsfig{file=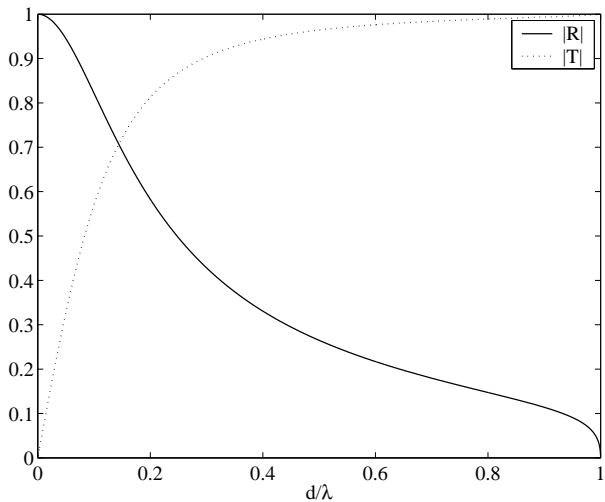, width=8cm} \caption{\emph{Reflection and transmission coefficient for a single
grid of unloaded wires as a function of the grid's spatial frequency.}} \label{RT1}
\end{figure}

Let the wires in the grid be first unloaded ($Z=0$). The reflection and transmission coefficients calculated
using eq.~(\ref{RT}) are depicted as a function of the normalized wire separation in Fig.~\ref{RT1}. We
qualitatively repeat the well known observation (e.g.~\cite{Yatsenko}) according to which at low frequencies the
grid is strongly reflective (the wave does not probe the fine details of the grid). Note that the normalized
frequency $d/\lambda=1$ corresponds to a certain kind of resonance \cite{Sergei, Yatsenko}:~at this frequency
the contribution from the other wires to the local electric field acting on the surface of the reference wire
adds up in phase. However, the field amplitude at the wire surface must be limited which is only possible when
the current induced to the grid is zero, thus, the grid is ideally transparent at this frequency.

For the calculation of the mesoscopic permittivity we need to choose the unit cell thickness~$s$. One rather
intuitive choice is $s=2r_0$. In addition to this, we partly follow the analysis of \cite{Schwartz,Guida} where
the the authors use an artificial interface shift $d/2$ on both sides of stacks of wire grids when representing
the actual structures as slabs made of homogeneous material.\footnote{In \cite{Schwartz, Guida} the lattice
constant is the same in both transversal (with respect to the wire axis) directions.} Thus, in the second
scenario $s=d$ (when representing the grid as a homogeneous slab we imagine that there is an artificial
interface shift $d/2$ on both sides of the grid).

\begin{figure}[t!]
\centering \epsfig{file=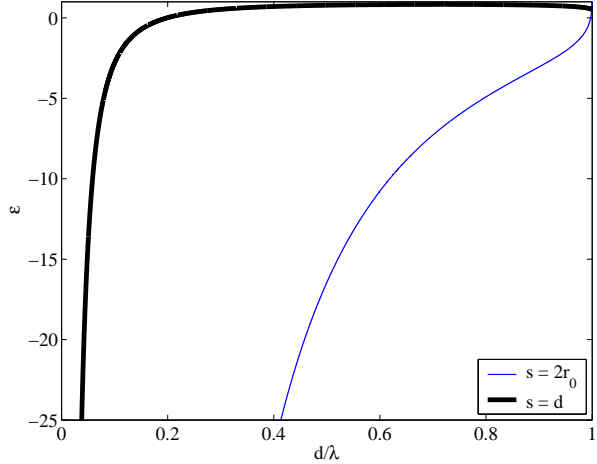, width=8cm} \caption{\emph{Mesoscopic effective permittivity of a single grid
of unloaded wires as a function of the grid's spatial frequency.}} \label{eeff1}
\end{figure}

The permittivities for the both aforementioned situations, calculated using eq.~(\ref{epseff1}), are shown in
Fig.~\ref{eeff1}. We notice that in both cases the permittivity is purely real as it should be since the wires
(and the background medium) are lossless. In the case $s=2r_0$ the frequency derivative of permittivity is
positive over the entire frequency range, thus, the permittivity function strictly satisfies the passivity
requirement. Interestingly, however, this is not the case when we introduce the artificial interface shift: the
permittivity has a maximum roughly at $d/\lambda=0.72$, and the slope becomes negative as the frequency
approaches the point $d/\lambda=1$. At these frequencies the permittivity function will lead to a physically
wrong sign for the imaginary part when inserted into the Kramers-Kroning relations. However, also the wire
spacing is comparable to the wavelength at these frequencies and the validity of the effective medium
description is weak. At frequencies below $d/\lambda\approx0.72$ we can, however, at least qualitatively make
some interesting observations from the results depicted in Fig.~\ref{eeff1}.

Typically in the quasi-static regime the effective permittivity of infinite lattices of unloaded wires (when
excited with normally incident planes waves) is described using the lossless Drude model
(e.g.~\cite{Pavel,Maslovski}):
\begin{equation}
\epsilon = 1 - \frac{\omega_0^2}{\omega^2},
\end{equation}
where $\omega_0$ is the angular ``plasma'' frequency which is determined by the structural parameters of the
grid. If we assume that instead of a single grid of wires we would investigate the properties of an infinite
wire lattice with rectangular periodicity (the lattice constant in both transversal directions would be $d$),
the normalized wire medium plasma frequency, estimated using equations available in \cite{Pavel, Maslovski},
would be $d/\lambda_0=0.20$. When $s=d$ we estimate from Fig.~\ref{eeff1} that $d/\lambda_0\approx0.20$.
Actually, a similar observation has been made already in \cite{Guida}: the authors observed that in case of an
interface shift corresponding roughly to $d/2$ the effective permittivity of an equivalent homogeneous slab
simulating the single grid closely approximates the permittivity of an infinite lattice. This is intuitively
rather clear after simple symmetry considerations \cite{Maslovski}. When the unit cell thickness $s=2r_0$ the
behavior of the permittivity function is dramatically different compared to the case $s=d$.

\begin{figure}[t!]
\centering \epsfig{file=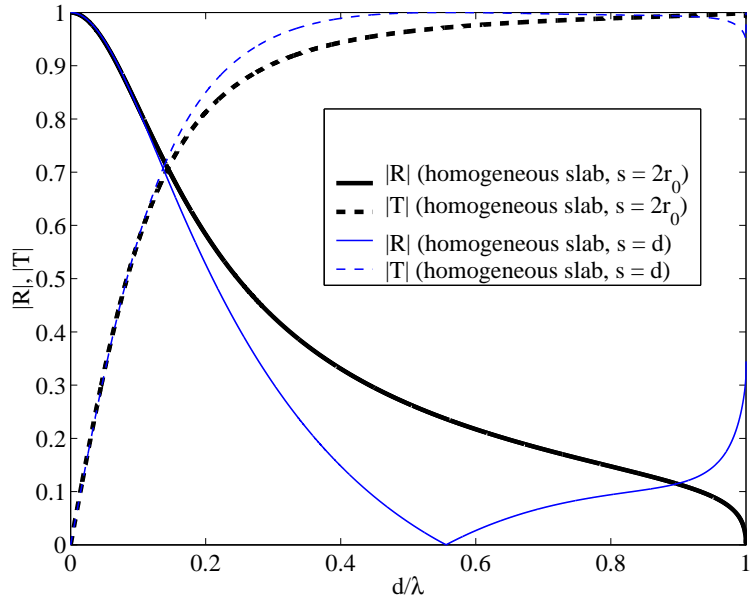, width=10cm} \caption{\emph{Reflection and transmission data calculated for the
homogenized slabs simulating a grid of unloaded wires.}} \label{RTc1}
\end{figure}

Next, we represent the actual wire grid as a slab of homogeneous material having thickness (in $x$-direction,
Fig.~\ref{grid}) $2r_0$ or $d$. The corresponding effective permittivities characterizing these homogeneous
slabs are those shown in Fig.~\ref{eeff1}. We use standard equations for calculating the reflection and
transmission coefficients through such homogeneous slabs, see e.g.~the transmission matrix formalism presented
in [22, Ch.~2.4]. The results are depicted in Fig.~\ref{RTc1}.

In the first case the effective slab thickness equals to the diameter of the wires, and we observe a very close
agreement between the exact result depicted in Fig.~\ref{RT1}, and the result calculated for the homogeneous
slab. For this specific example we have, thus, verified that the effective permittivity depicted in
Fig.~\ref{eeff1} correctly describes the actual (microscopic) polarization response of the grid, and also allows
to reproduce the reflection and transmission results with a normal plane-wave incidence.

When the slab thickness equals the wire spacing the agreement between the exact result and the result depicted
in Fig.~\ref{RTc1} is good roughly up to $d/\lambda=0.1$. At higher frequencies the discrepancy is visible
(please remember also the discussion concerning the meaningfulness of the permittivity function for this case at
frequencies above $d/\lambda=0.72$). It is evident from the results that the effective permittivity for this
situation (see Fig.~\ref{eeff1}) does not allow to reproduce the actual reflection and transmission results with
normal plane wave incidence. Obviously the difficulties arise from the correct definition of the equivalent
homogeneous slab thickness. However, permittivity defined when $s=d$ gives a good estimate for the macroscopic
polarization response of an infinite wire lattice (at least up to frequencies $d/\lambda=0.72$, in our example
case).

\subsubsection{Capacitively loaded wires}

\begin{figure}[t!]
\centering \epsfig{file=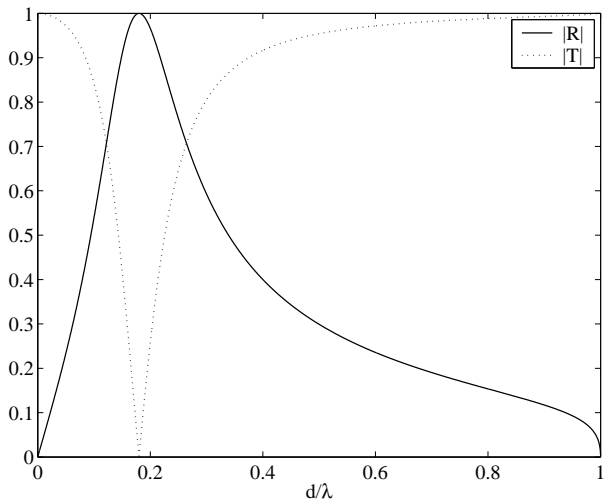, width=8cm} \caption{\emph{Reflection and transmission coefficient for a single
grid of wires loaded with bulk capacitors as a function of the grid's spatial frequency.}} \label{RT12}
\end{figure}

Let us next load the wires by bulk capacitors whose value is chosen to be $C=1$ pF. It is known \cite{Pavel_PRE}
that when the load capacitance value is infinitely small, the dispersion characteristics of a lattice of such
wires resemble the characteristics of a lattice of point dipole scatterers. When a realistic load capacitance
value and the insertion period are correctly chosen, this equivalence holds still, to some extend, at low
frequencies \cite{Pavel_PRE}. Thus, one can consider a properly capacitively loaded wire grid as a grid of short
wire (dipole) scatterers. Non-physical dispersion results (defined through reflection and transmission) for
grids of such inclusions have been presented earlier e.g.~in \cite{Dolling, Zhou}.

The reflection and transmission coefficients calculated using eq.~(\ref{RT}) are shown in Fig.~\ref{RT12}. We
see that there appears a stop-band not observed in the previous case (grid of unloaded wires). The behavior seen
in Fig.~\ref{RT12} can be explained (qualitatively) by regarding the grid as a bulk load, consisting of a series
connection of inductance and capacitance, in a waveguide representing free space wave propagation \cite{Sergei}.
At low frequencies the impedance of the grid is strongly capacitive and the load represents an open circuit in
the waveguide, thus, waves can propagate through the grid. At some frequency the wires are in self-resonance,
and in this case the waveguide is effectively short-circuited leading to a stop-band \cite{Pavel_PRE, Sergei}.

\begin{figure*}[t!]
\centerline{\subfigure[{\emph{Permittivity over a wide frequency
range.}}]{\includegraphics[width=7.5cm]{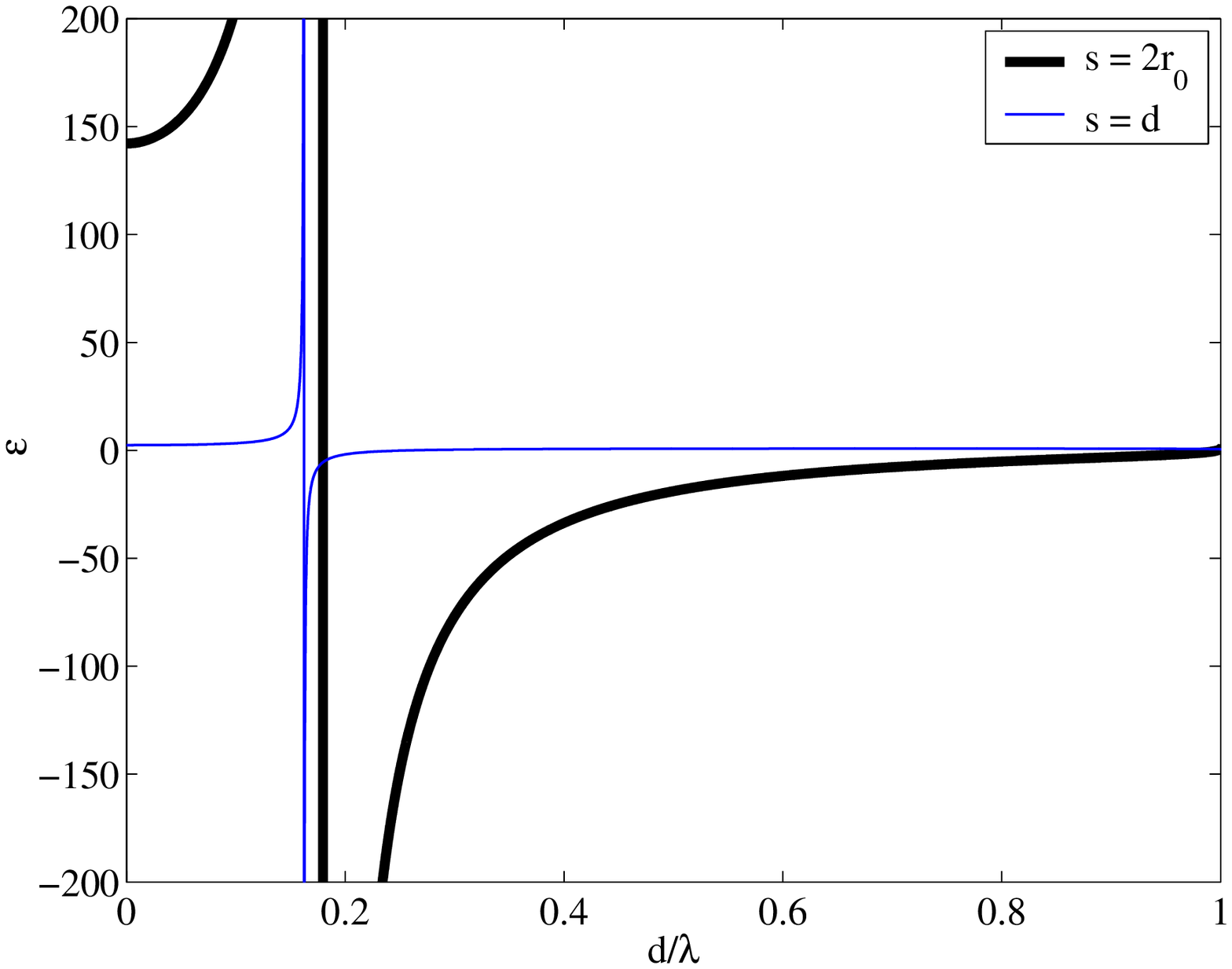} \label{eeff12}} \hfil \subfigure[{\emph{The same result as in
(a), but plotted over a more narrow frequency range.}}]{\includegraphics[width=7.5cm]{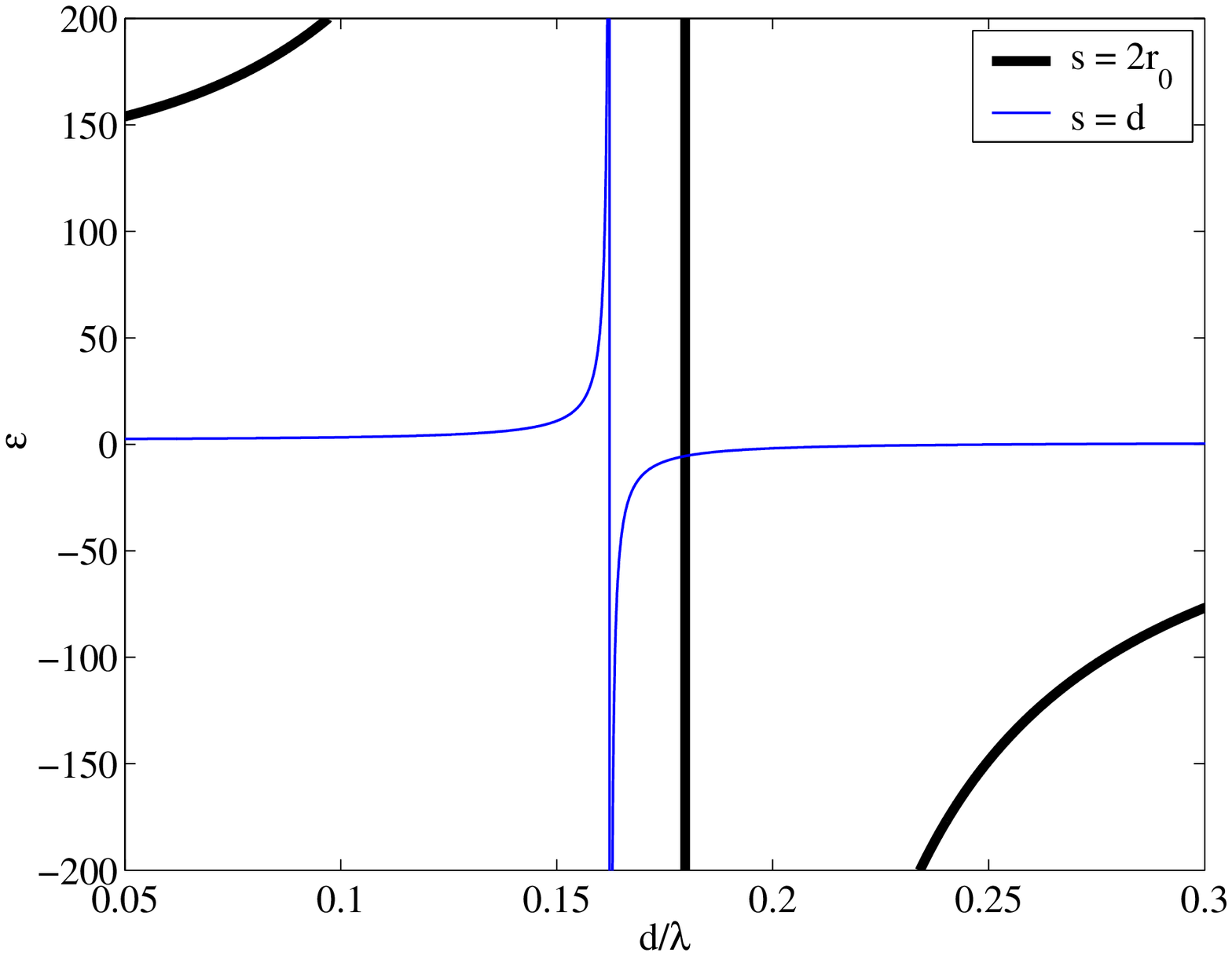}
\label{eeff12_2}}} \caption{{\emph{Mesoscopic effective permittivity of a single grid of wires loaded with bulk
capacitors as a function of the grid's spatial frequency.}}} \label{eeff12_t}
\end{figure*}

The mesoscopic effective permittivity calculated using eq.~(\ref{epseff1}) for this case is shown in
Fig.~\ref{eeff12_t}. The permittivity is purely real, and we observe that it obeys a Lorentz-type behavior,
which is consistent with the effective permittivity deduced for a semi-infinite capacitively loaded wire medium
\cite{Pavel_PRE}:
\begin{equation}
\epsilon = 1 + \frac{\Lambda\omega_0^2}{\omega_0^2-\omega^2},
\end{equation}
where $\Lambda$ is an amplitude factor depending on the load capacitance value and the structural parameters of
the grid (also $\omega_0$ in this case depends on both the structural parameters and the load capacitance
value). This behavior is physically due to the fact that in the self-resonance the distributed impedance value
on the wire surfaces is low and allows strong current to be induced to the wire. Strong induced current directly
means strong polarization response, therefore the very high (resonant) value of the permittivity. Moreover, the
choice of the unit cell thickness $s$ again dictates the strength of the polarization response. Please note that
in reality, due to finite conductivity of the wires and parasitic losses in the lumped elements, the effective
permittivity would show a significant imaginary part in the vicinity of the resonance as is imposed by the
Kramers-Kroning relations. Similarly as with the grid of unloaded wires we observe that the permittivity, when
defined using $s=2r_0$, is a monotonically growing function outside the resonant band. A closer investigation of
the permittivity function, defined when $s=d$, reveals that the slope of the function is negative between the
frequencies $d/\lambda=0.72$ ... $d/\lambda=1$. The behavior at these frequencies is therefore similarly
nonphysical as with the grid of unloaded wires (however, also the effective medium description is weak).

Authors of \cite{Maslovski} considered an infinite lattice of capacitively loaded wires and deduced the
following quasi-static limit for the effective permittivity in case of relatively high capacitive impedance (the
notation is clear from Fig.~\ref{grid}):
\begin{equation}
\epsilon^{\rm s} = 1 + \frac{Cl}{\epsilon_0sd}. \label{qlim}
\end{equation}
When the unit cell thickness is $s=2r_0$ and $s=d$, eq.~(\ref{qlim}) predicts $\epsilon^{\rm s}=142$ and
$\epsilon^{\rm s}=2.4$, and the the corresponding values read from Fig.~\ref{eeff12} are exactly the same. For
an infinite lattice of capacitively loaded wires with $d$ as the lattice constant in both transversal directions
and the structural parameters considered here, the normalized plasma frequency is roughly $d/\lambda_0=0.168$
\cite{Pavel_PRE}. From Fig.~\ref{eeff12_t} we see that the corresponding frequencies are $d/\lambda_0=0.180$
($s=2r_0$) and $d/\lambda_0=0.163$ ($s=d$). Thus, again the introduction of the artificial interface shift
allows us to rather well approximate the refractive properties of an infinite lattice (at least in terms of the
effective plasma frequency and static effective permittivity).

\begin{figure*}[t!]
\centerline{\subfigure[{\emph{Unit cell thickness $s=2r_0$.}}]{\includegraphics[width=7.5cm]{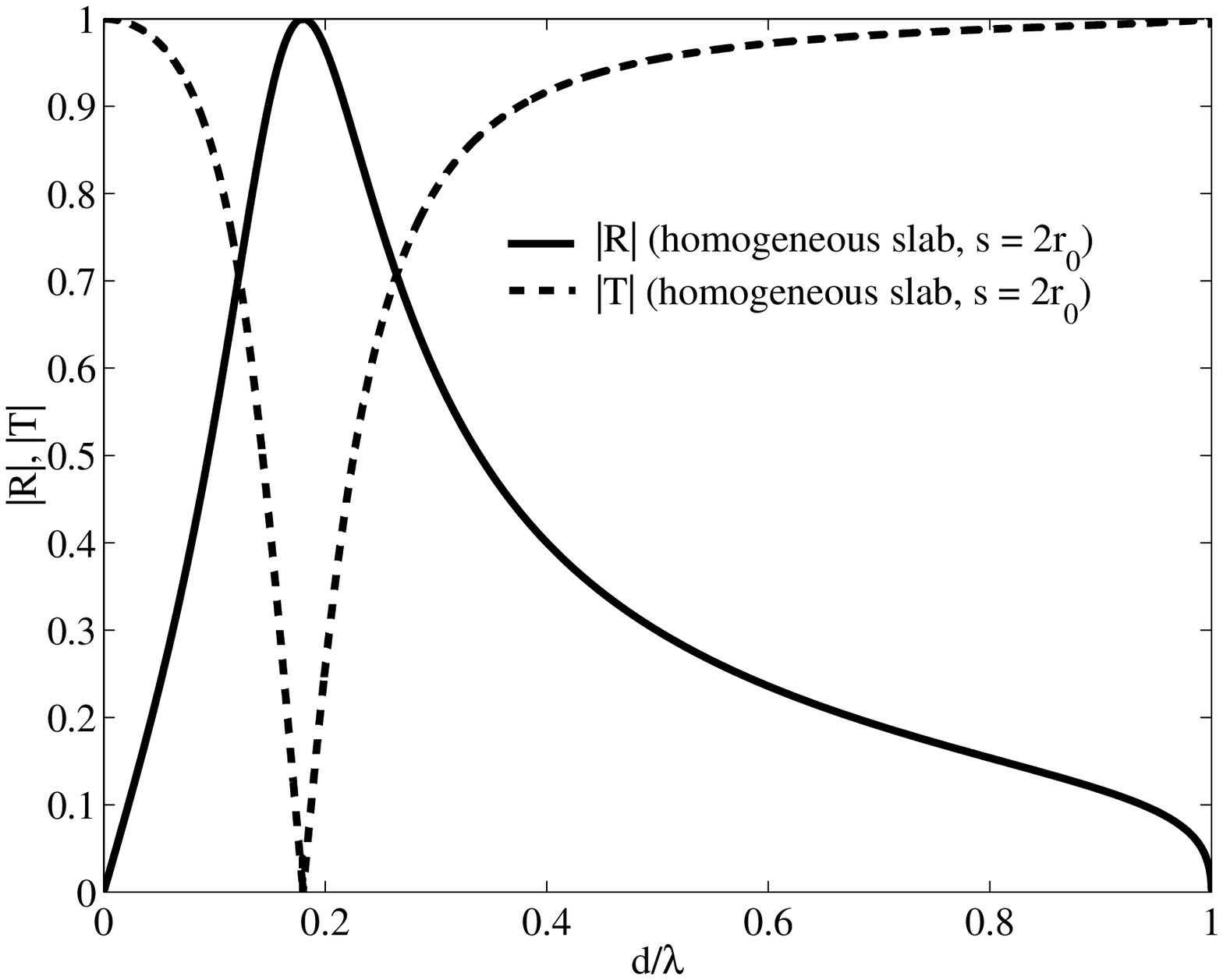}
\label{RTc12_a}} \hfil \subfigure[{\emph{Unit cell thickness
$s=d$.}}]{\includegraphics[width=7.5cm]{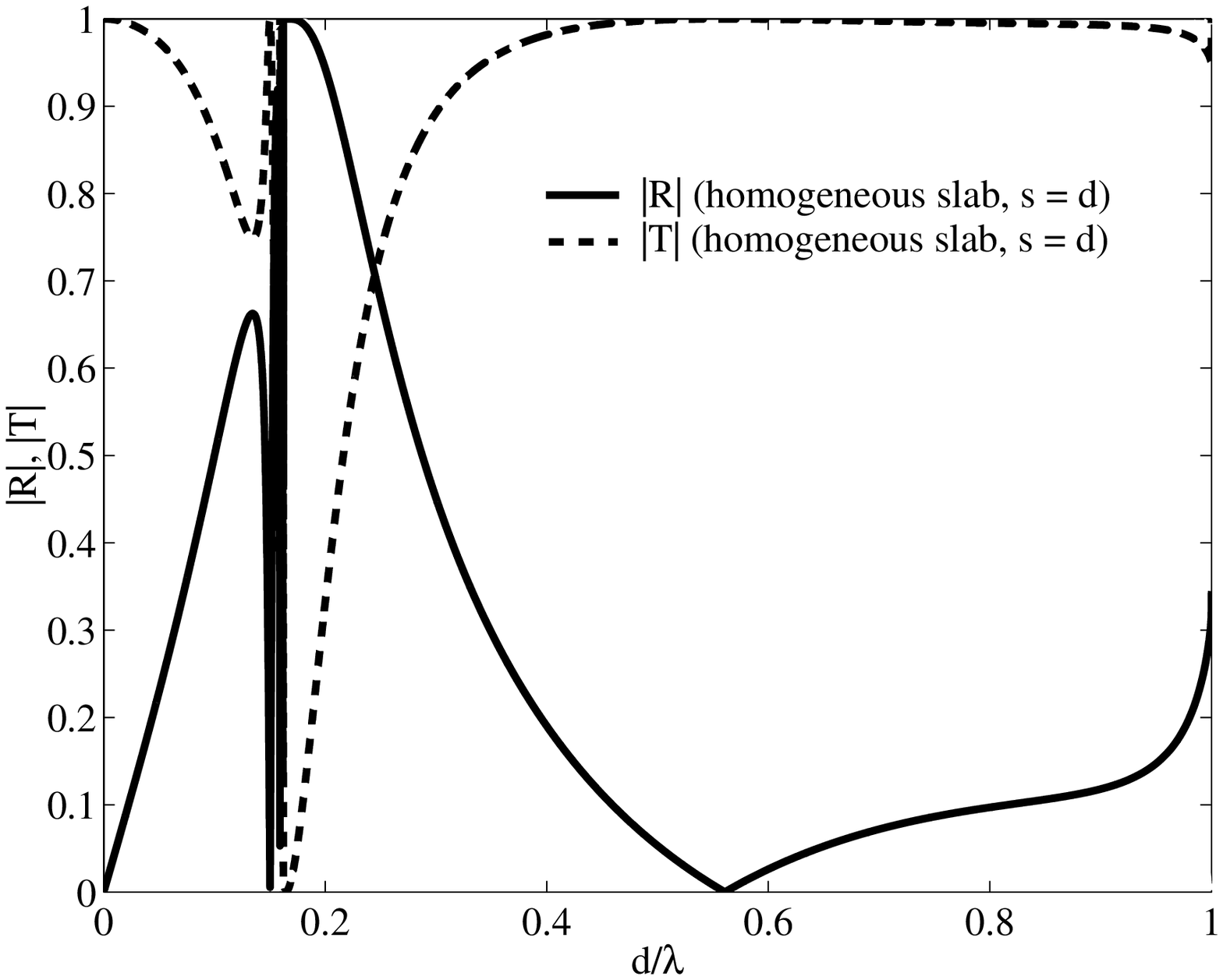} \label{RTc12_b}}} \caption{{\emph{Reflection and
transmission data calculated for the homogenized slabs simulating a grid of capacitively loaded wires.}}}
\label{RTc12}
\end{figure*}

The reflection and transmission coefficients calculated for the homogeneous slabs are shown in Fig.~\ref{RTc12}.
We observe that when the unit cell thickness equals the wire diameter the mesoscopic effective permittivity,
depicted in Fig.~\ref{eeff12_t}, represents the grid correctly both from the polarization and
reflection/transmission points of view. If an artificial increase in the slab thickness is introduced the
reflective/transmissive properties dramatically differ from the exact results close to the self-resonant
frequency (as with the grid of unloaded wires the results agree well up to $d/\lambda=0.1$). Near that
self-resonance the absolute value of the effective refractive index is very large, thus, the electrical slab
length is very large. In a slab whose physical thickness noticeably differs from zero such condition can be
predicted to cause thickness resonances that would be seen as sharp transmission maxima. This, indeed seems to
be the case according to Fig.~\ref{RTc12}.


\subsubsection{Wires loaded with a parallel connection of capacitance and inductance}

\begin{figure}[b!]
\centering \epsfig{file=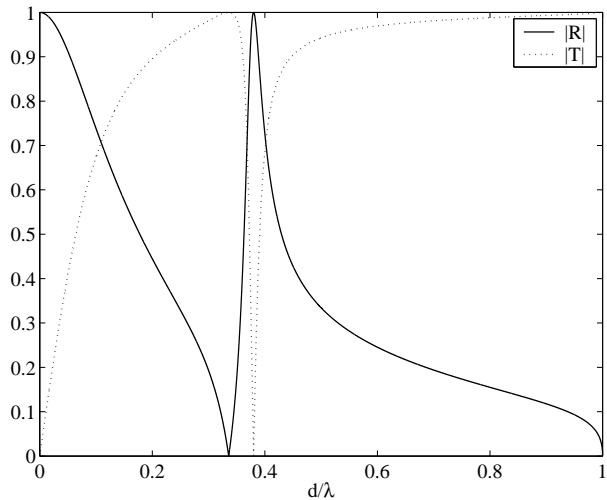, width=8cm} \caption{\emph{Reflection and transmission coefficient for a single
grid of wires loaded with a parallel connection of bulk capacitors and inductors as a function of the grid's
spatial frequency.}} \label{RT13}
\end{figure}

In this case the wires are loaded with bulk impedances consisting of a parallel connection of capacitance and
inductance. The capacitance value is kept the same as before, the load inductance value is $L=1$ nH, and there
appears a parallel resonance at $d/\lambda_{\rm l}=0.34$. The reflection and transmission coefficients for such
a loaded grid are depicted in Fig.~\ref{RT13}. Once again, the behavior can be straightforwardly explained using
the idea of a waveguide loaded (in parallel) with a bulk load. At low frequencies the the impedance of the load
is strongly inductive and the waveguide is effectively short circuited. When the load circuitry is in parallel
resonance the impedance of the load tends ideally to infinity, and waves can propagate through the grid. With
our load values and structural parameters a series resonance occurs just above the parallel load resonance. This
resonance is caused by the inductive wire segments (that can be modeled as bulk inductors) and load capacitors,
and is observed as the sharp reflection maximum in Fig.~\ref{RT13}.

\begin{figure*}[t!]
\centerline{\subfigure[{\emph{Permittivity over a wide frequency
range.}}]{\includegraphics[width=7.5cm]{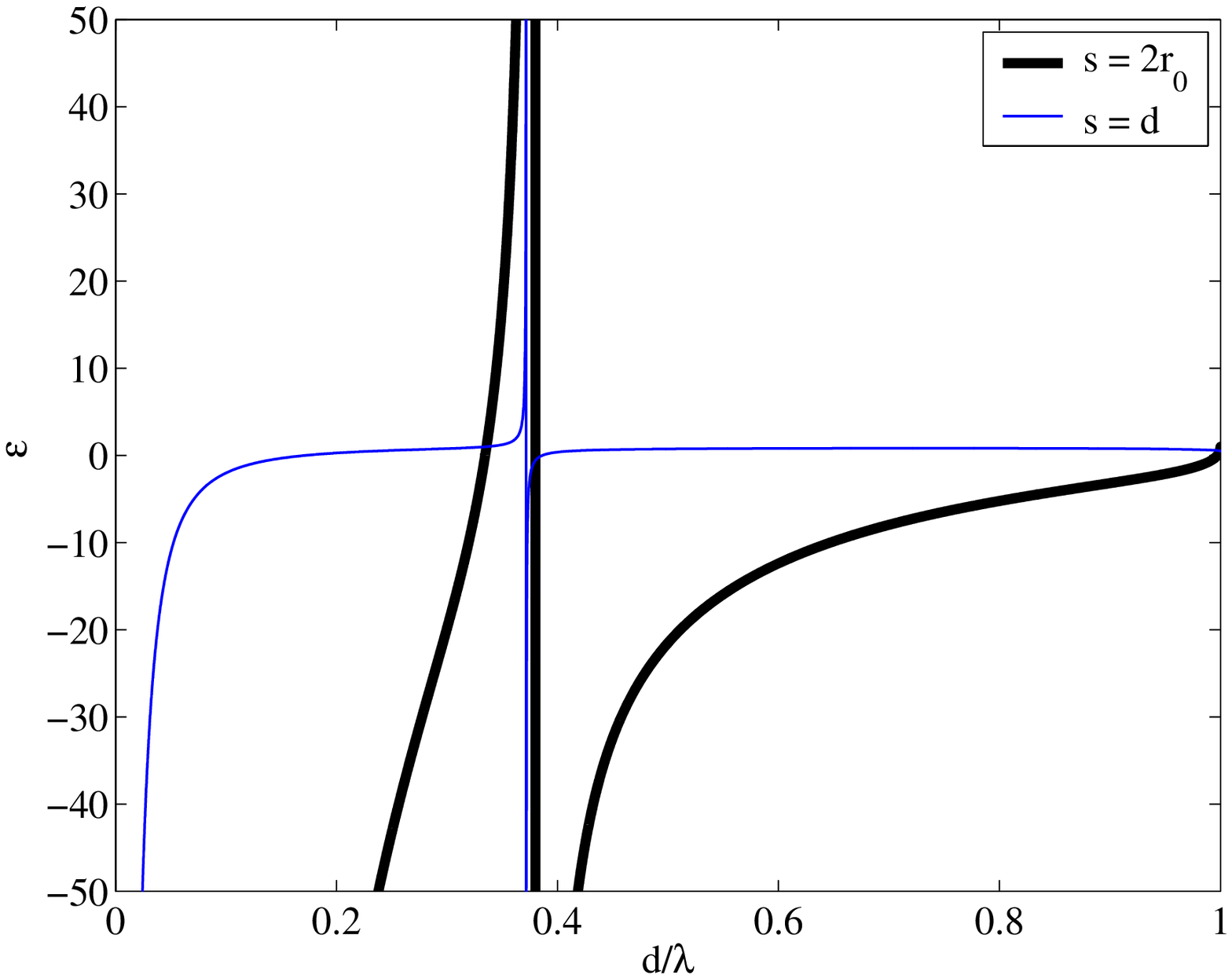} \label{eeff13}} \hfil \subfigure[{\emph{The same result as in
(a), but plotted over a more narrow frequency range.}}]{\includegraphics[width=7.5cm]{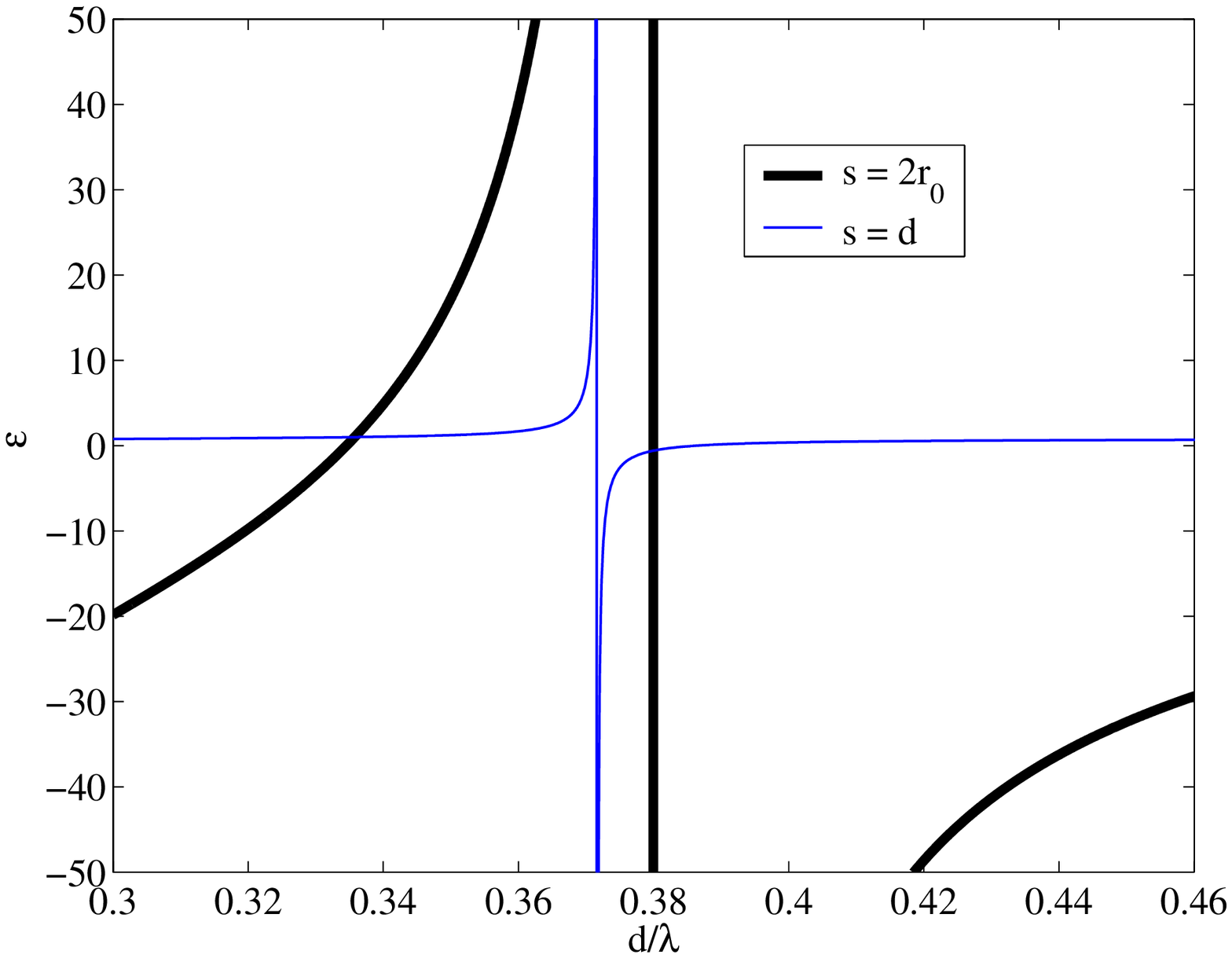}
\label{eeff13_2}}} \caption{{\emph{Mesoscopic effective permittivity for a grid of wires loaded with a parallel
connection bulk capacitors and inductors as a function of the grid's spatial frequency.}}} \label{eeff13_t}
\end{figure*}

\begin{figure*}[b!]
\centerline{\subfigure[{\emph{Unit cell thickness $s=2r_0$.}}]{\includegraphics[width=7.5cm]{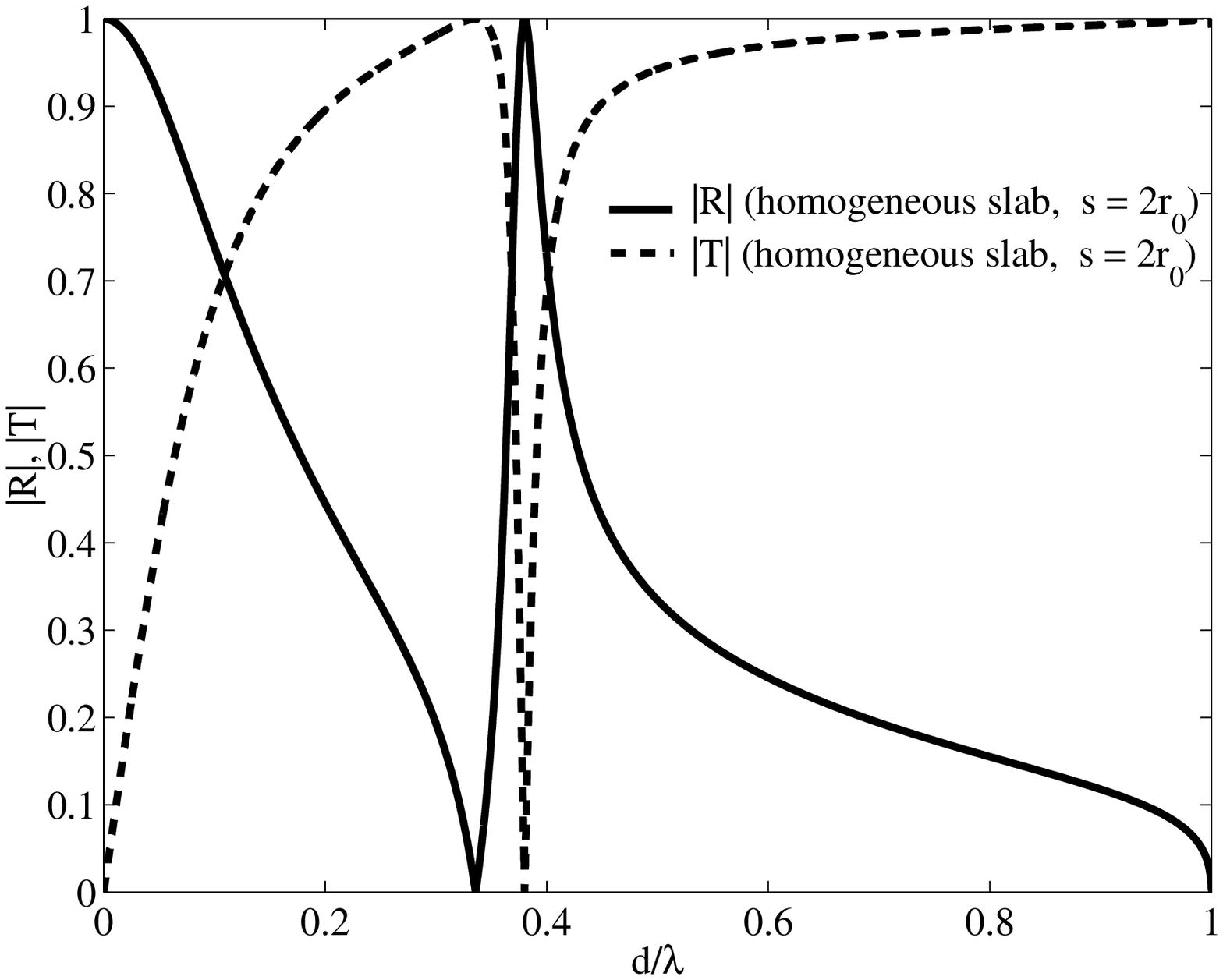}
\label{RTc13_a}} \hfil \subfigure[{\emph{Unit cell thickness
$s=d$.}}]{\includegraphics[width=7.5cm]{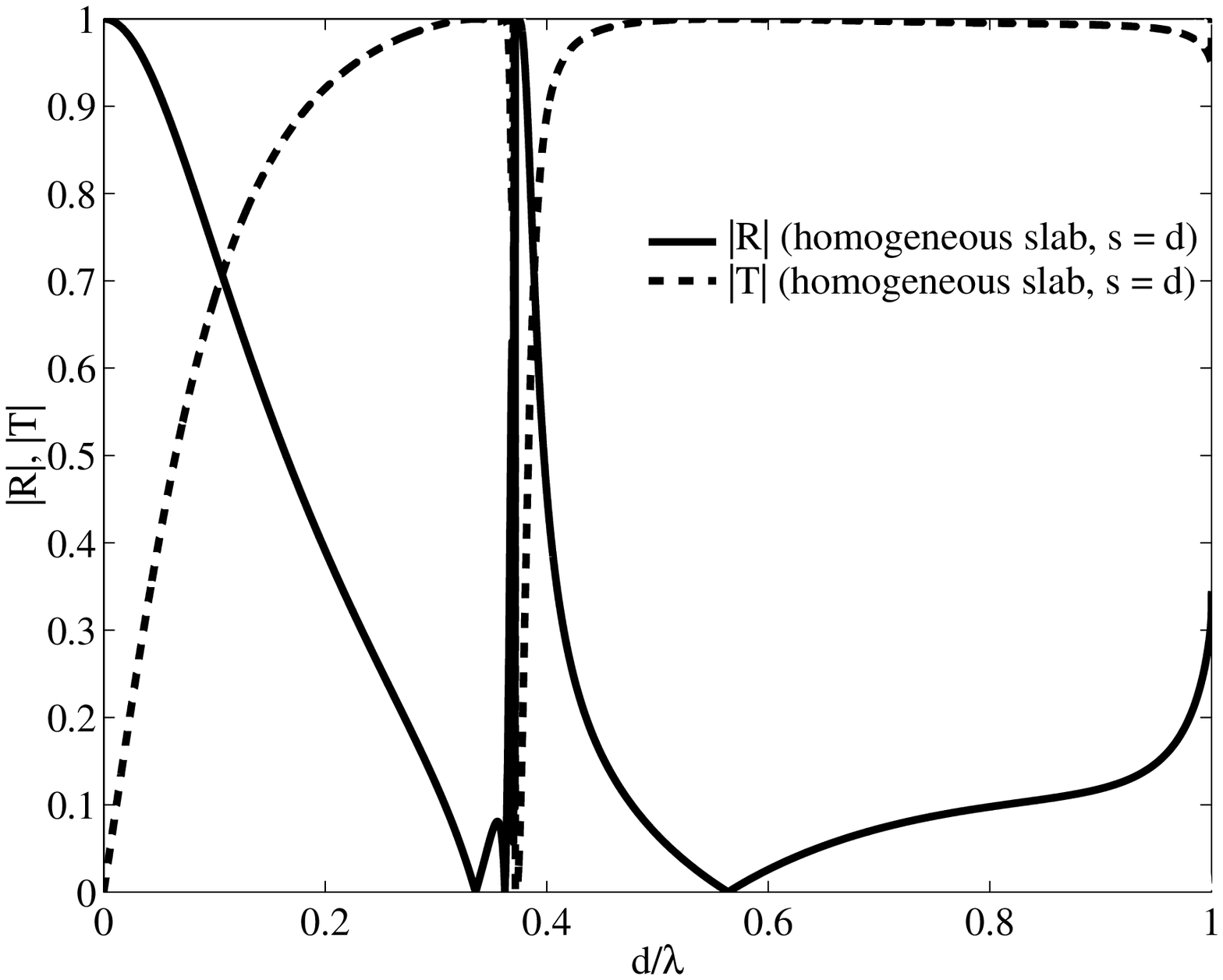} \label{RTc13_b}}} \caption{{\emph{Reflection and
transmission data calculated for the homogenized slabs simulating a grid of capacitively and inductively loaded
wires.}}} \label{RTc13}
\end{figure*}

When looking at the calculated mesoscopic effective permittivity, shown in Fig.~\ref{eeff13_t}, we observe that
the permittivity crosses unity at the parallel load resonance (this is seen as the crossing point of the
permittivity functions in Fig.~\ref{eeff13_t}). This explains also the total transmission at this frequency.
Around the frequency point of series load resonance the behavior is qualitatively the same as in the case of
capacitively loaded wires. Note again that the permittivity is purely real in both cases. However, again the
frequency derivative of $\epsilon$, defined using $s=d$, is negative at frequencies higher than
$d/\lambda=0.72$.


The reflection and transmission coefficients calculated for the homogeneous slabs are shown in Fig.~\ref{RTc13}.
The behavior is qualitatively the same as already described with the previous examples:~when the thickness of
the homogeneous slab equals the wire diameter, the slab correctly describes the actual grid both from the
polarization and reflection/transmission points of view.

In short intermediate conclusion related to the single grids, the mesoscopic effective permittivity, defined
directly through induced electric dipole moment density and averaged electric field, is a physically sound
function with all the loading scenarios over the spectral range where the effective medium description is valid.
Moreover, the calculated examples indicate that the grid can accurately be represented both for polarization
(equivalently, for refraction when defined through induced dipole moment density and averaged electric field)
and reflection/transmission as a homogeneous slab having the proposed permittivity when the unit cell thickness
equals the diameter of the wires. This observation holds when the diameter is very small. When an artificial
increase in the unit cell thickness corresponding to $d/2$ is introduced, the permittivity function representing
the homogeneous slab does not allow to reproduce the reflection/transmission results for the actual grid.
However, the function gives (up to a certain frequency) a rather good approximation for the effective
permittivity of an infinite lattice.


\subsection{Double grid of wires}

\subsubsection{Unloaded wires}

\begin{figure}[b!]
\centering \epsfig{file=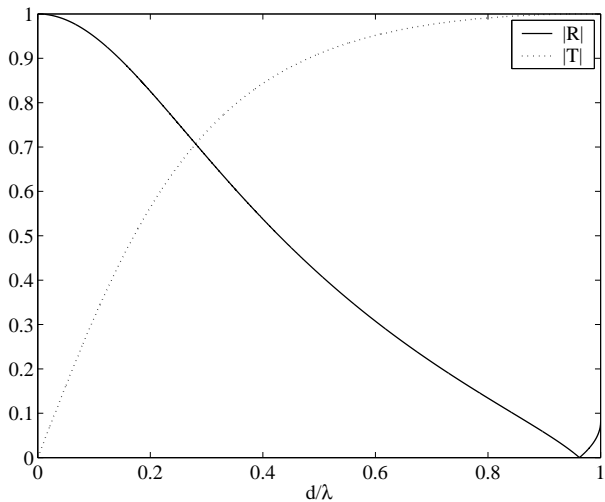, width=8cm} \caption{\emph{Reflection and transmission coefficient for a double
grid of unloaded wires as a function of the grid's spatial frequency.}} \label{RT21}
\end{figure}

The reflection and transmission coefficients (referred to the plane of the left-most grid in Fig.~\ref{2grid})
are shown in Fig.~\ref{RT21} (we have used eq.~(\ref{RT}) by also taking into account the scattered field
created by the second grid, and the phase shift between the grids). The coefficients qualitatively resemble the
results depicted in Fig.~\ref{RT1} for a single grid of unloaded wires with the exception that the zero point of
reflection coefficient has shifted to a lower frequency due to the influence of the second grid.

\begin{figure*}[t!]
\centerline{\subfigure[{\emph{Material parameters over a wide frequency
range.}}]{\includegraphics[width=7.5cm]{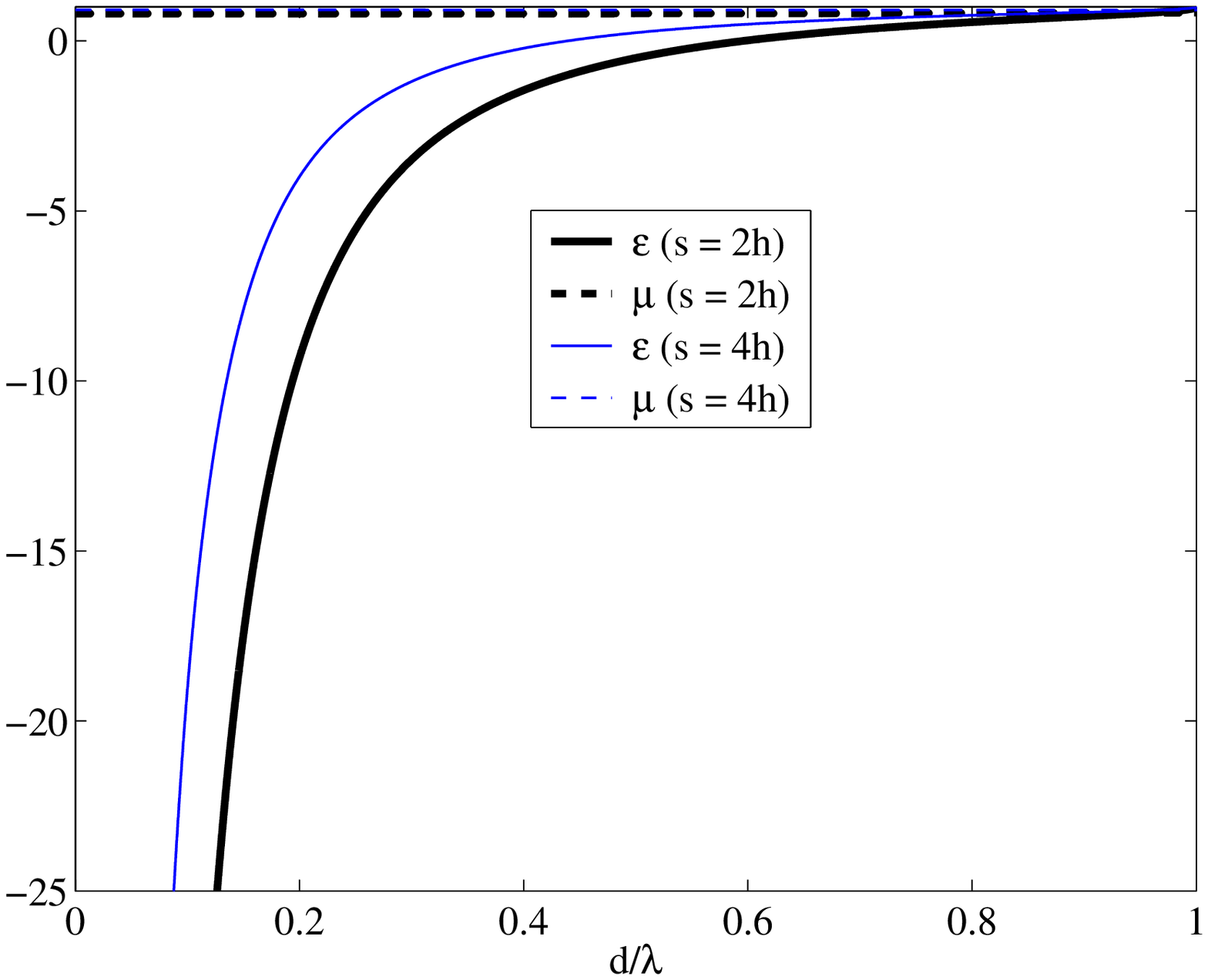} \label{eeff2}} \hfil \subfigure[{\emph{The same result as in
(a), but plotted over a more narrow amplitude range.}}]{\includegraphics[width=7.5cm]{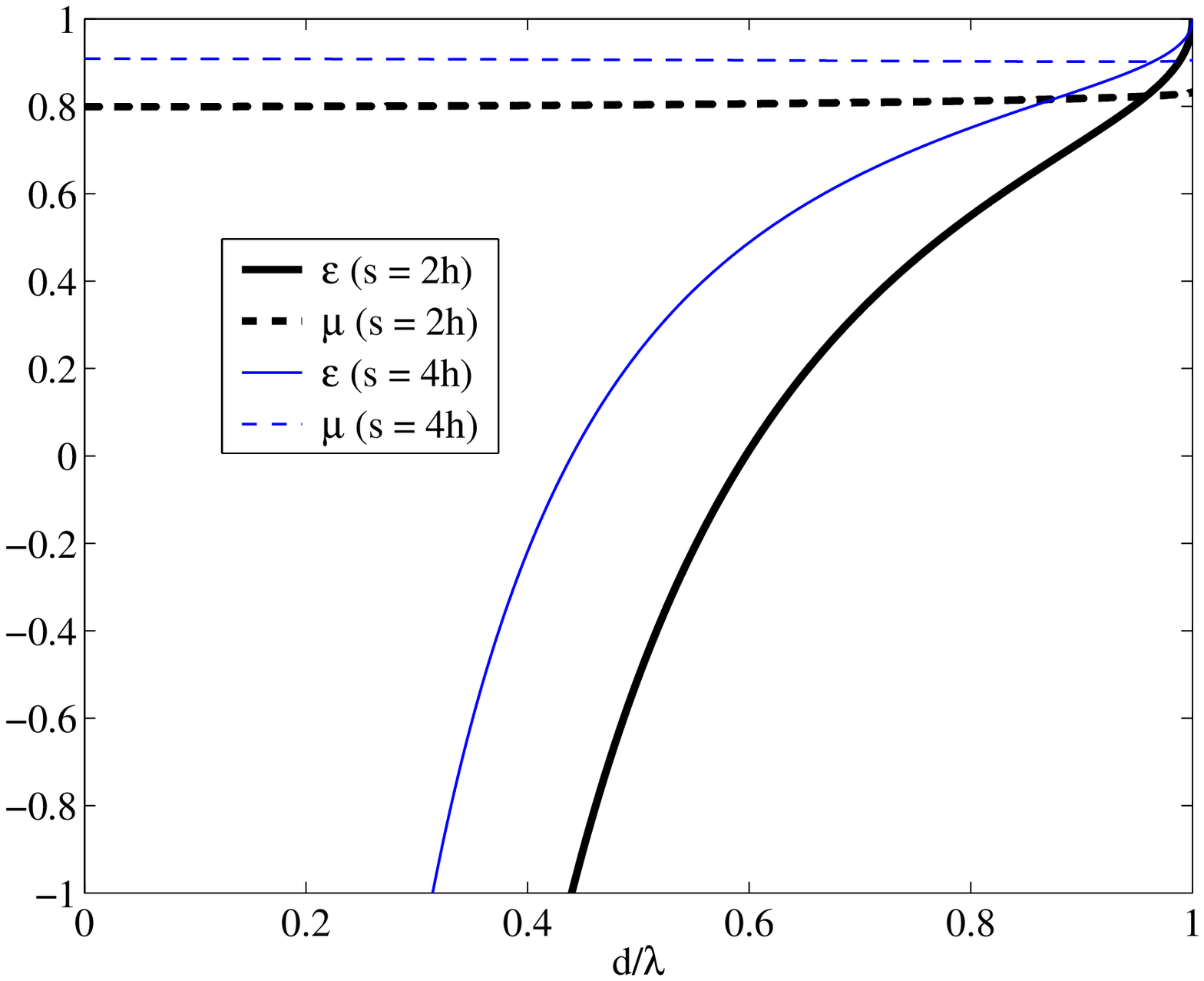}
\label{eeff2_2}}} \caption{{\emph{Mesoscopic effective permittivity and permeability of a double grid of
unloaded wires as a function of grid's spatial frequency.}}} \label{eeff2_t}
\end{figure*}

The mesoscopic effective material parameters of the equivalent homogeneous slab are shown in Fig.~\ref{eeff2_t}.
In the double grid analysis we set the unit cell thickness equal to $s=2h$ (accurately corresponding to the
actual physical thickness of the structure) and $s=4h$, following the analysis of \cite{Schwartz, Guida}. Please
note that in this case the lattice constants in the transversal directions (with respect to the wire axis) are
different, and we set the artificial interface shift equal to half of the lattice constant in $y$-direction
(Fig.~\ref{2grid}).

\begin{figure}[t!]
\centering \epsfig{file=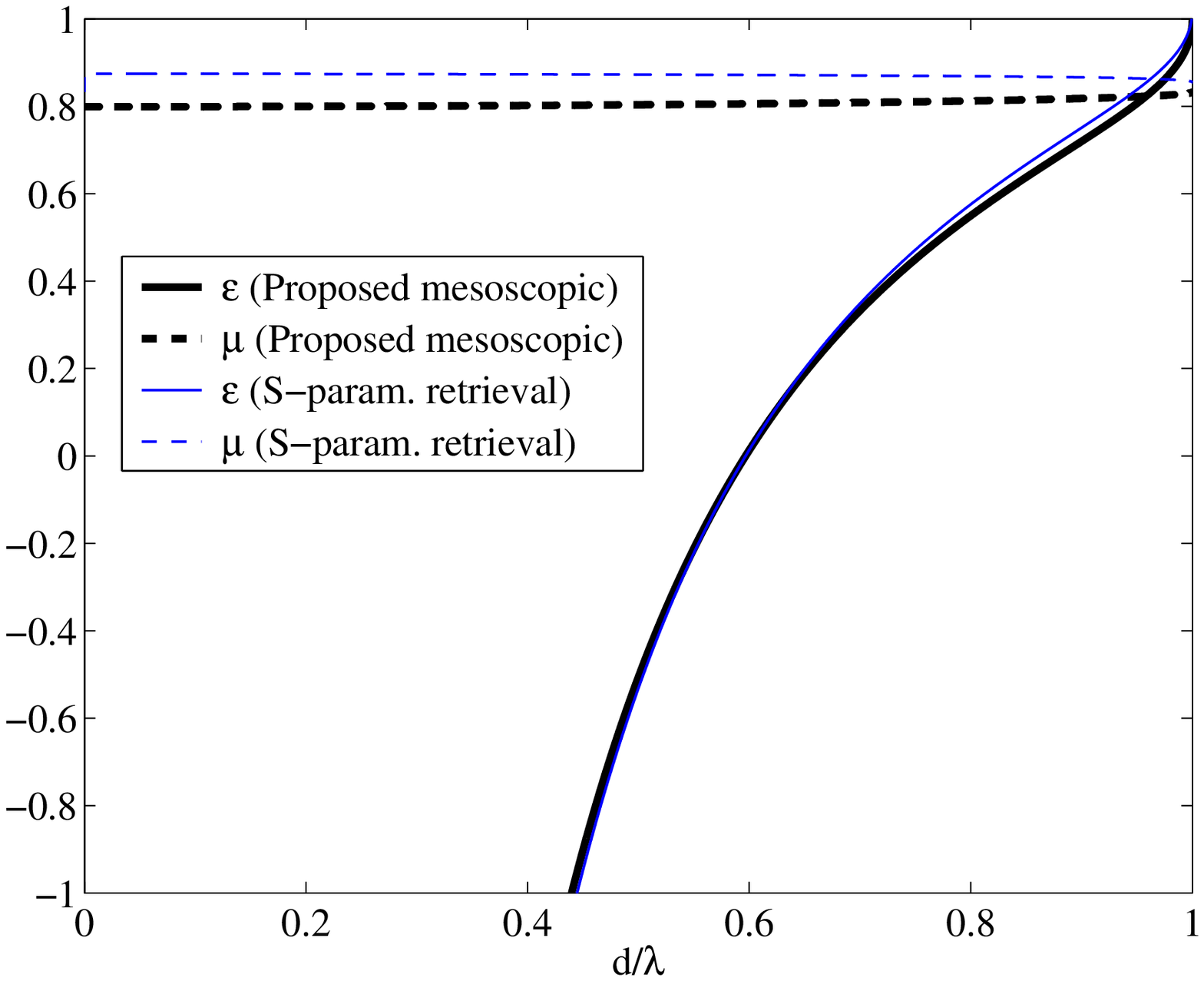, width=10cm} \caption{\emph{Comparison between the mesoscopic effective
material parameters obtained using the proposed method and the $S$-parameter retrieval method.}} \label{U_comp}
\end{figure}

\begin{figure}[b!]
\centering \epsfig{file=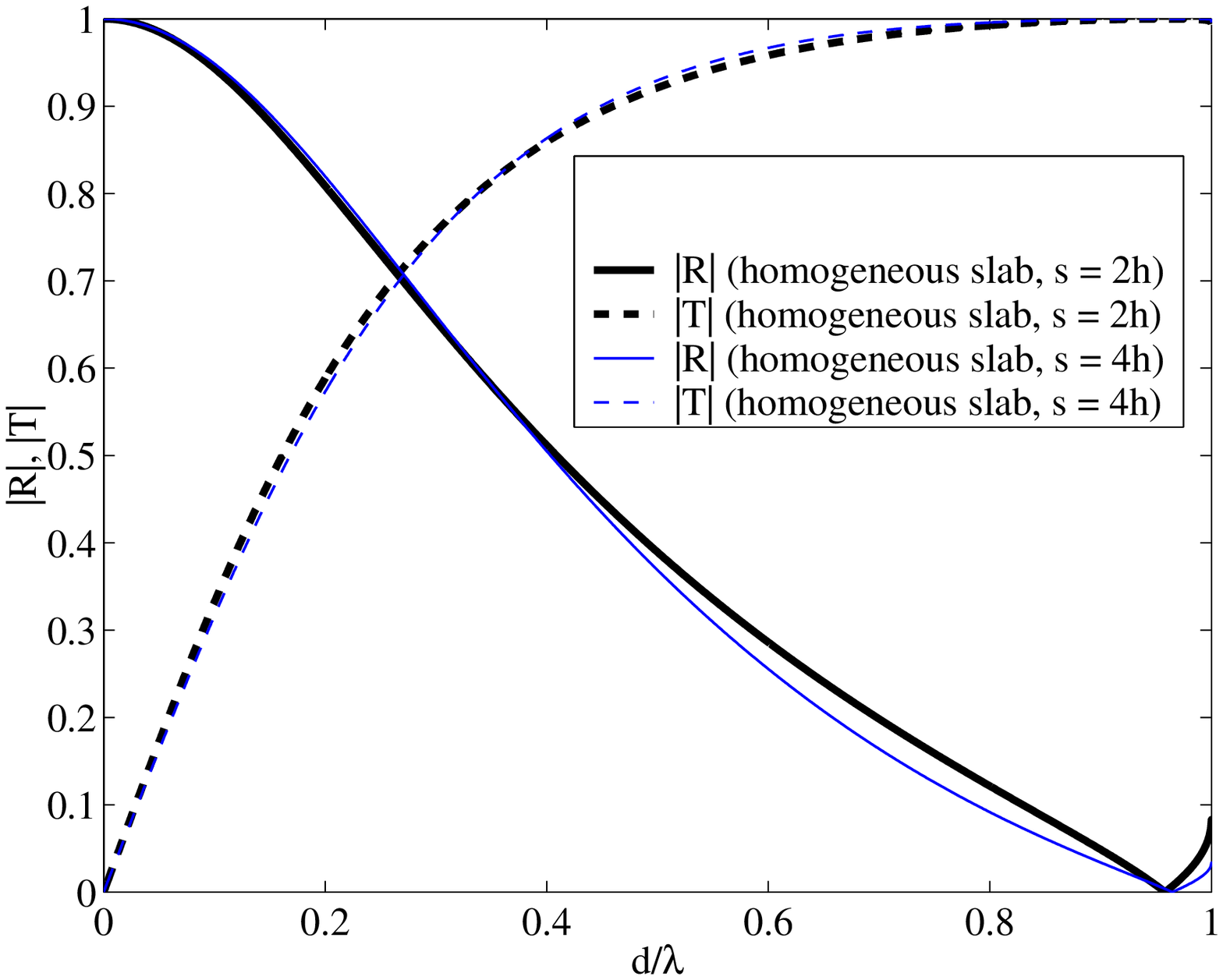, width=10cm} \caption{\emph{Reflection and transmission data calculated for the
homogenized slabs simulating a double grid of unloaded wires.}} \label{RT2c1}
\end{figure}

We observe that all the material parameters are purely real, and the frequency derivative of every parameter is
a growing function over the entire spectral range. The effective permittivity behaves qualitatively in a similar
manner as introduced in connection with the single grid. However, in this case the double grid is very slightly
diamagnetic:~the static effective permeabilities for the two homogeneous slabs, defined from Fig.~\ref{eeff2_2},
are 0.80 ($s=2h$) and 0.91 ($s=4h$). A similar observation related to diamagnetism has often been observed in
connection with the results obtained using the $S$-parameter retrieval method \cite{David}. For an infinite
lattice having lattice constant $2h$ in $x$-direction and $d$ in $y$-direction, equations available in
\cite{Pavel_PRE} predict that the normalized plasma frequency of such a wire medium is $d/\lambda_0=0.50$. The
normalized plasma frequencies observed in Fig.~\ref{eeff2_t} correspond roughly to $d/\lambda_0\approx0.60$
($s=2h$) and $d/\lambda_0\approx0.44$ ($s = 4h$).

Fig.~\ref{U_comp} depicts a comparison between the effective material parameters obtained using the proposed
method and the $S$-parameter retrieval method (for the $S$-parameter retrieval method we have used equations
available in \cite{Smith1}). In this example the thickness of the equivalent homogeneous slab equals the
physical thickness ($s=2h$), and we observe that the two methods give rather closely agreeing results. However,
the frequency derivative of permeability function, obtained using the $S$-parameter retrieval procedure, is very
slightly negative over the entire plotted frequency range. The static effective permeability obtained using the
$S$-parameter retrieval procedure equals 0.87.

Fig.~\ref{RT2c1} shows the reflection and transmission coefficients for the homogeneous slabs characterized by
the mesoscopic effective parameters depicted in Fig.~\ref{eeff2_t}. We see that with both slab thicknesses the
results agree rather closely with the exact results depicted in Fig.~\ref{RT21}. Additional calculations (not
shown) indicate that when the interface shift is increased the disagreement between the reflection/transmission
results for the actual structure and for the homogeneous slab increases, and eventually there appears a null in
the reflection coefficient around $d/\lambda\approx 0.5$.

\subsubsection{Capacitively loaded wires}

The reflection and transmission coefficients for this case are depicted in Fig.~\ref{RT12}. Once again the
behavior resembles the behavior of a single grid of capacitively loaded wires with the important exception that
there appears a narrow transmission window inside the wider stopband. This transmission window originates
physically from the fact that the currents in the grids cancel each other at this frequency (when the phase
shift between the grids is taken into account, remember that the reflection coefficient is referred to the plane
of the leftmost grid in Fig.~\ref{2grid}).

\begin{figure}[b!]
\centering \epsfig{file=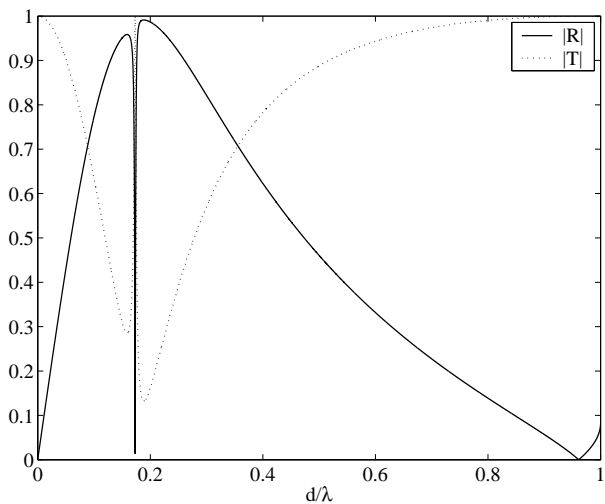, width=8cm} \caption{\emph{Reflection and transmission coefficient for a double
grid of wires loaded with bulk capacitors as a function of the grid's spatial frequency.}} \label{RT12}
\end{figure}
\begin{figure*}[t!]
\centerline{\subfigure[{\emph{Material parameters over a wide frequency
range.}}]{\includegraphics[width=7.5cm]{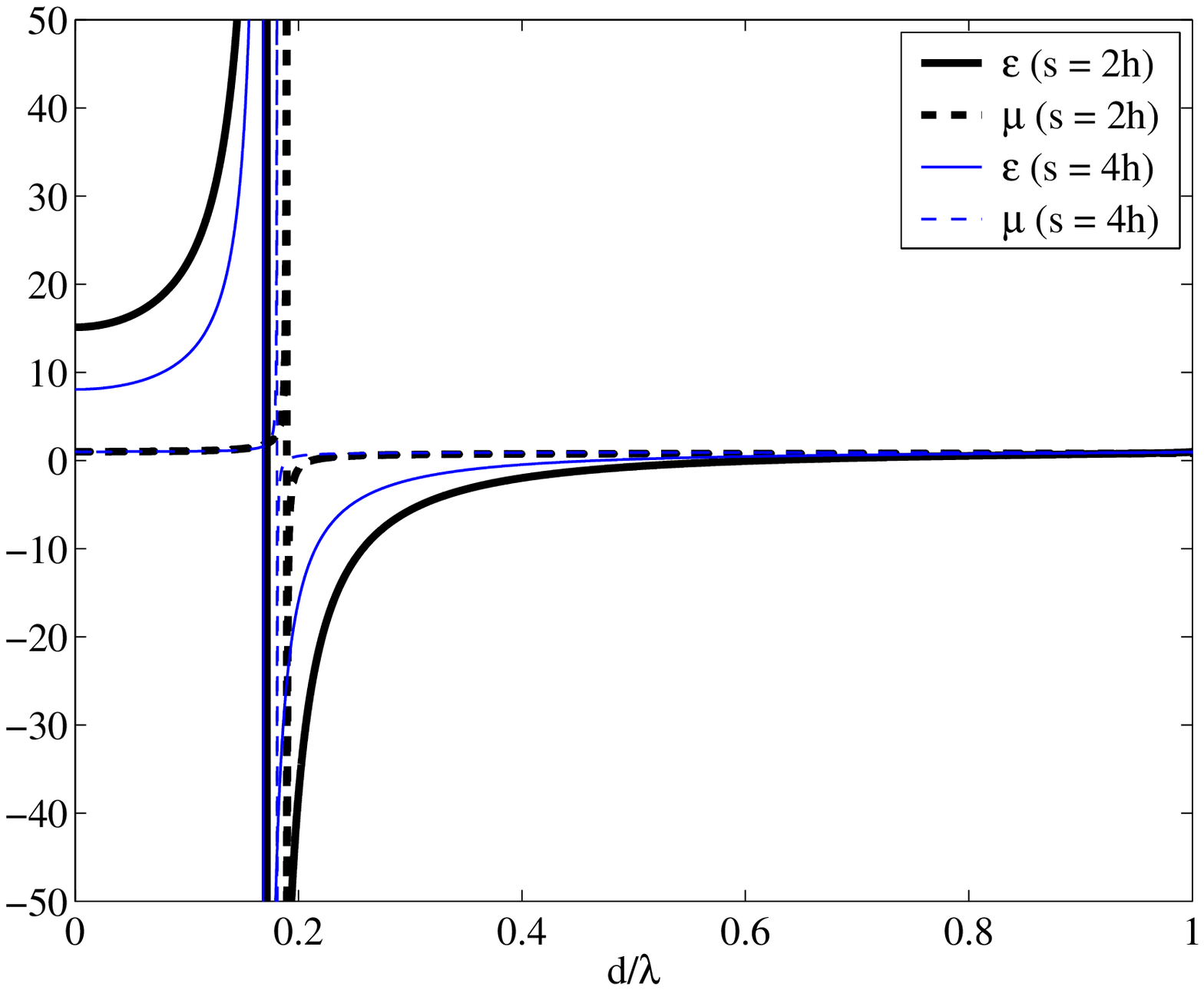} \label{eeff22}} \hfil \subfigure[{\emph{The same result as in
(a), but plotted over a more narrow frequency range.}}]{\includegraphics[width=7.5cm]{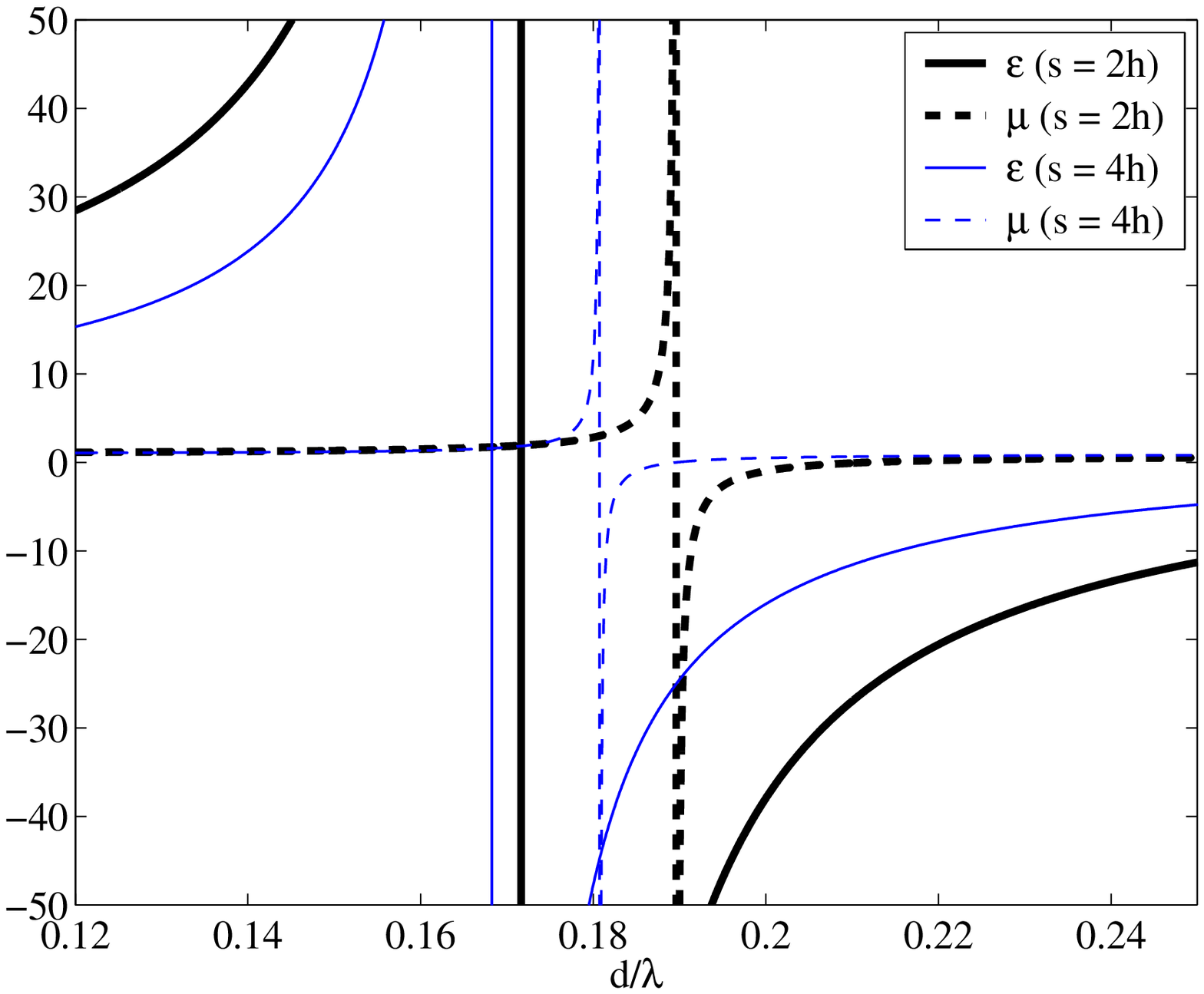}
\label{eeff22_2}}} \caption{{\emph{Mesoscopic effective permittivity and permeability of a double grid of wires
loaded with bulk capacitors as a function of the grid's spatial frequency.}}} \label{eeff22_t}
\end{figure*}

\begin{figure*}[b!]
\centerline{\subfigure[{\emph{Comparing the mesoscopic effective
permittivities.}}]{\includegraphics[width=7.5cm]{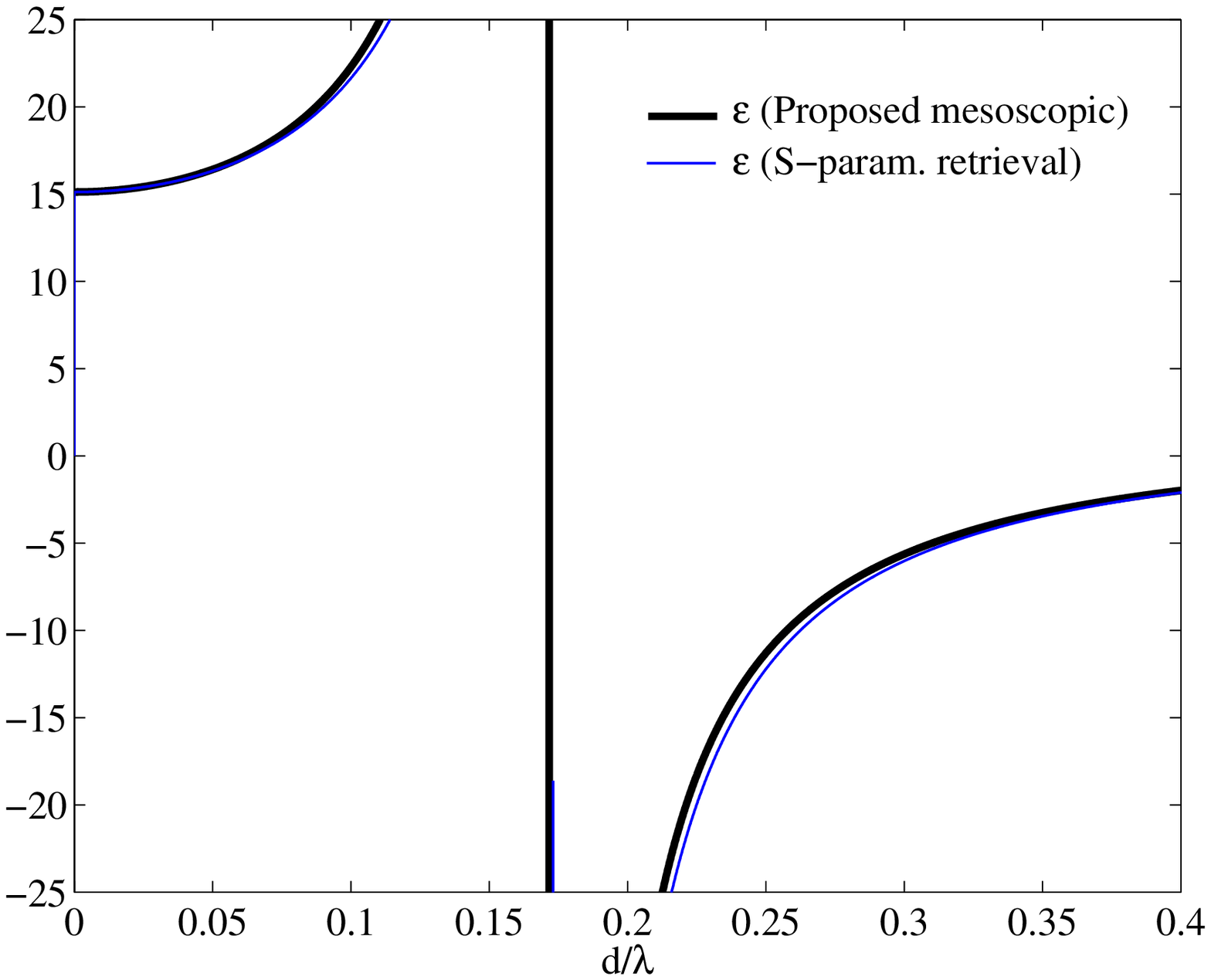} \label{C_compe}} \hfil \subfigure[{\emph{Comparing
the mesoscopic effective permeabilities.}}]{\includegraphics[width=7.5cm]{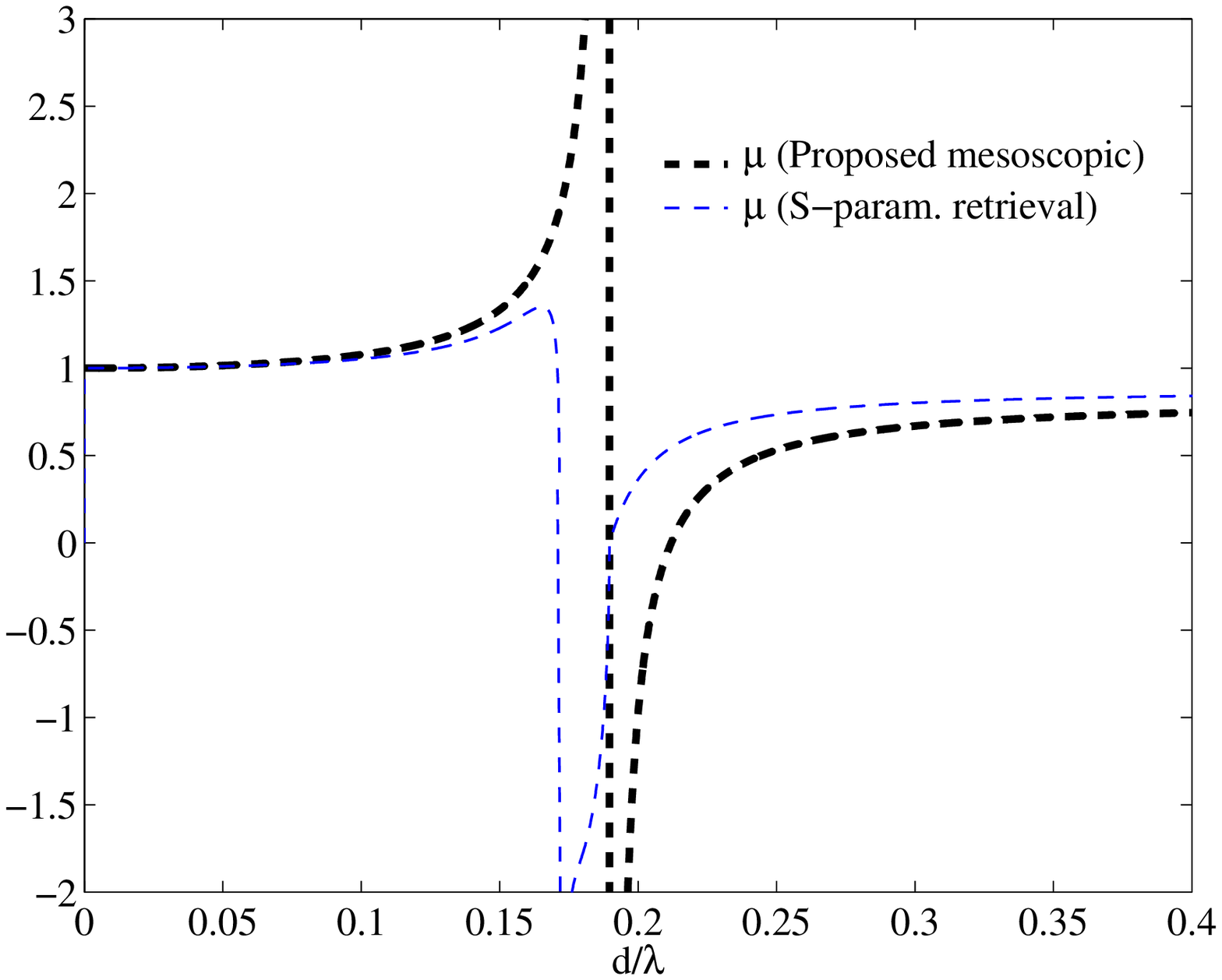} \label{C_compm}}}
\caption{{\emph{Comparison between the mesoscopic effective material parameters obtained using the proposed
method and the $S$-parameter retrieval method.}}} \label{C_comp}
\end{figure*}

The calculated mesoscopic effective material parameters are depicted in Fig.~\ref{eeff22_t}. Please note that
all the material parameters are purely real and behave in a physically sound manner (the frequency derivative is
positive everywhere outside the resonance region). Moreover, the frequency dependence of both parameters clearly
resembles the behavior predicted by the classical Lorentz model. The behavior in the effective permittivity
arises from the self-resonance of the wires, i.e.~when the load circuitry is in series resonance the current
induced to the wires will be strong leading to a strong polarization response. The resonant permeability arises
from circulating out-of-phase currents flowing in the grids. The difference in the current amplitudes flowing in
the grids is minimized at a slightly higher frequency as compared to the deep reflection minimum observed in
Fig.~\ref{RT12}. Therefore the reflection minimum (in this particular case) does not exactly correspond to the
resonant point of permeability. It is also worth to observe that resonance in permeability is a very narrowband
phenomenon compared to the resonance in permittivity. This is because the mutual cancelation of the grid
currents occurs in a narrow frequency window due to very strong mutual coupling (very small grid separation)
between the grids.

It is also worthwhile to notice that both mesoscopic effective parameters are negative in a narrow frequency
band just above the resonance of permeability, thus, the polarization response at this frequency range can be
described by assigning ``negative effective refractive index'' for the structure. We have to bear in mind,
however, the important physical limitations outlined above for the introduction of effective parameters for thin
complex slabs. One should also observe that the amplitudes of permittivity and permeability
(Fig.~\ref{eeff22_t}) are at the same (negative) level only over an extremely narrow frequency band. This means
that the structure (with these particular structural dimensions and load values) would be very difficult to
match to incident radiation. When the unit cell thickness is $s = 2h$ and $s=4h$, eq.~(\ref{qlim}) predicts
$\epsilon^{\rm s}=7.7$ and $\epsilon^{\rm s}=4.5$. The corresponding values read from Fig.~\ref{eeff22_t} are
$\epsilon^{\rm s}=15.1$ and $\epsilon^{\rm s}=8.1$. Static permeabilities for both effective slab thicknesses
equal unity.


\begin{figure*}[b!]
\centerline{\subfigure[{\emph{Unit cell thickness $s=2h$.}}]{\includegraphics[width=7.5cm]{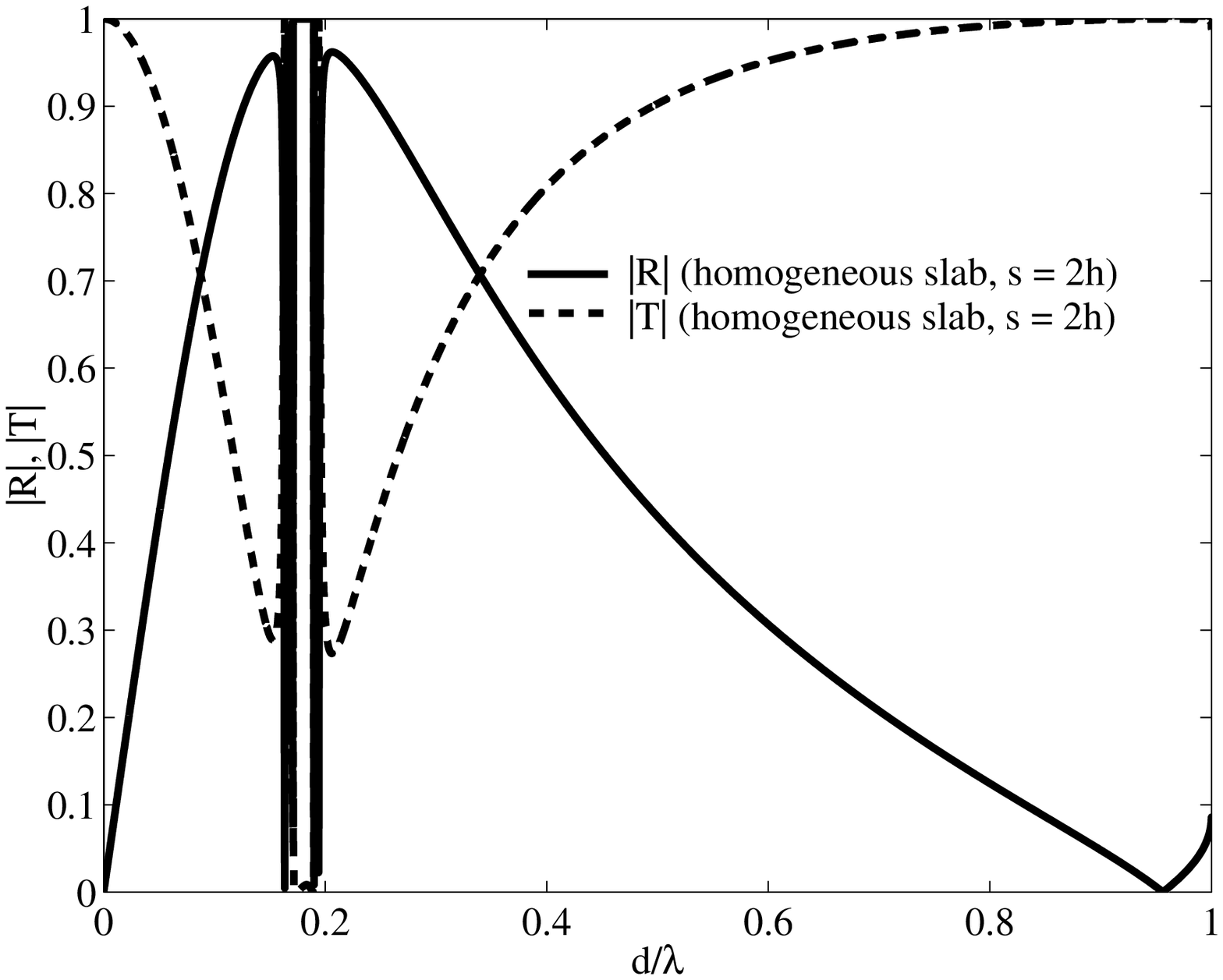}
\label{RT2c21_a}} \hfil \subfigure[{\emph{Unit cell thickness
$s=4h$.}}]{\includegraphics[width=7.5cm]{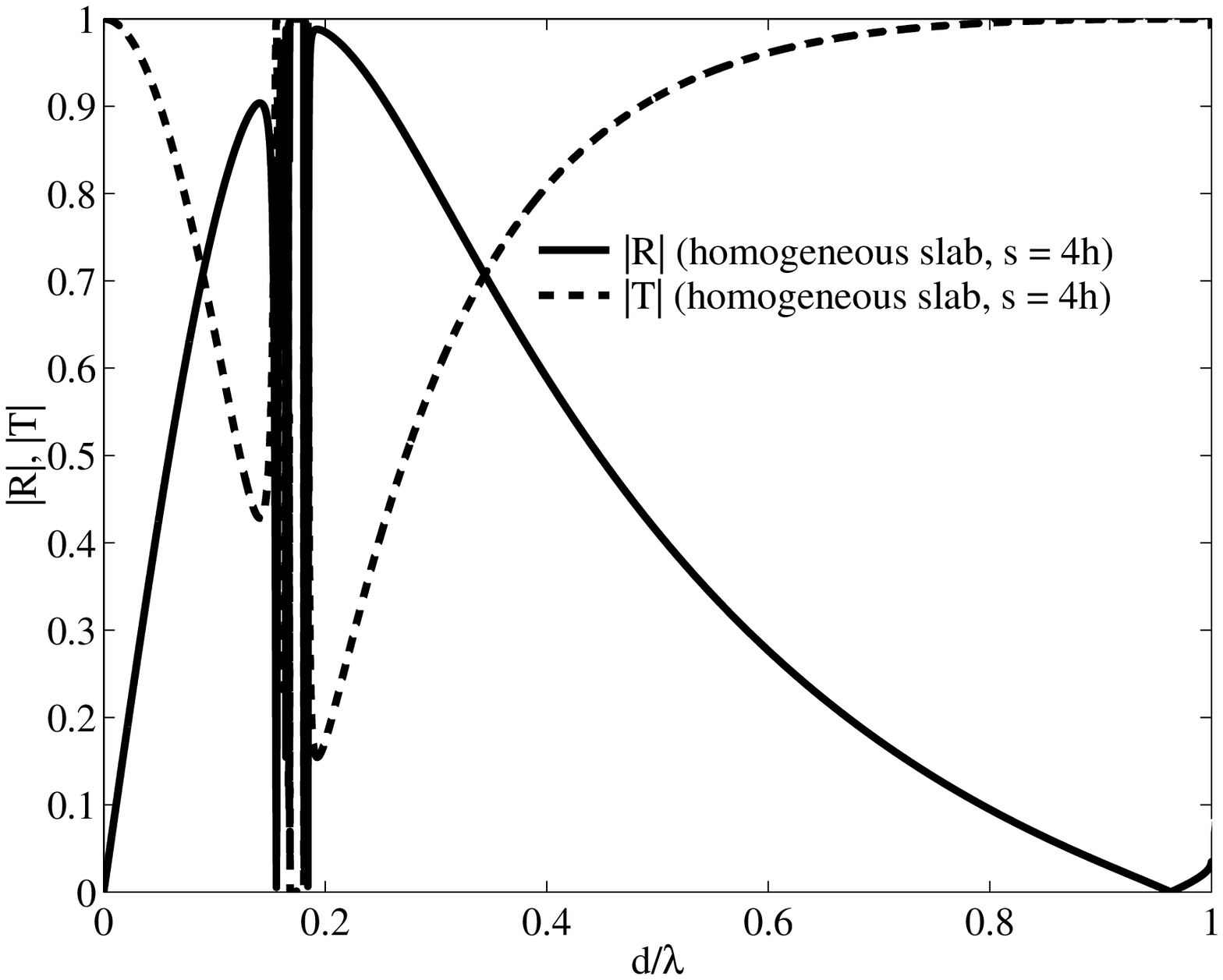} \label{RT2c21_b}}} \caption{{\emph{Reflection and
transmission data calculated for the homogenized slabs simulating a double grid of capacitively loaded wires.}}}
\label{RT2c21}
\end{figure*}

Comparison between the results obtained using the proposed method and the $S$-parameter retrieval method is
depicted in Fig.~\ref{C_comp}. The thickness of the equivalent homogeneous slab equals the physical thickness
($s=2h$). Interestingly, the effective permittivities agree rather well over the whole depicted frequency range
(visible disagreement is observed only close to the permittivity resonance point). However, the effective
permeabilities obtained using different methods behave in a visibly different manner:~the predicted magnetic
response of the double grid is much weaker according to the $S$-parameter retrieval results. Also, it is evident
that the function predicted by the $S$-parameter retrieval method cannot be represented using the causal Lorentz
model.

The reflection and transmission coefficients, calculated when the actual structure is represented as a slab of
homogeneous material (the material parameters are those depicted in Fig.~\ref{eeff22_t}), are shown in
Fig.~\ref{RT2c21}. We notice that in the vicinity of the resonant point of material parameters there appears
several thickness resonances. At other frequencies the agreement between the exact result (Fig.~\ref{RT12}) and
the results calculated for the homogeneous slabs is rather good.

We can conclude that the effective material parameters, shown in Fig.~\ref{eeff22_t}, correctly describe what
kind of polarization mechanism takes place in the microstructural level. However, when the actual grid structure
is represented as a homogenized slab, the proposed mesoscopic parameters do not allow to exactly reproduce the
reflection/transmission results. Visible disagreement is observed close to the resonant points of the material
parameters. Close to these resonant frequencies the electrical slab thickness is very large, and this leads to
several closely located thickness resonances not observed with the actual grid structure. Note that when the
wires are non-resonant, e.g.~unloaded as in the previous subchapter, the agreement between the
reflection/transmission results for the actual structure and homogenized slabs is rather good over the entire
spectral range. This indicates that when the slab is electrically very dense it is difficult to assign the
correct equivalent slab thickness reproducing the scattering scenario (intuitively this is clear since the
double-grid cannot physically correspond to a sample of any homogeneous material). It it important to bear in
mind that the above mentioned deficiencies are avoided with the $S$-parameter retrieval method since the method
is based on the fulfillment of the actual scattering scenario. However, as is shown by the above example (and
will further be shown below), the material parameter functions obtained using the $S$-parameter retrieval
results might not satisfy the causality or the passivity requirement (see also e.g.~\cite{Linden, Dolling,
Shalaev, Zhou, Koschny}).

\subsubsection{Wires loaded with a parallel connection of capacitance and inductance}

\begin{figure}[b!]
\centering \epsfig{file=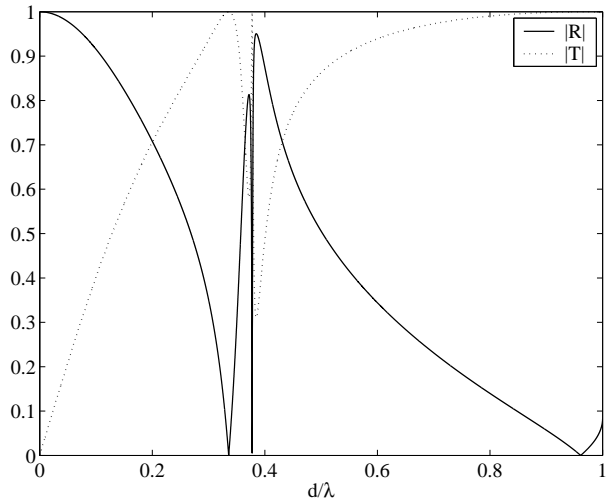, width=8cm} \caption{\emph{Reflection and transmission coefficient for a double
grid of wires loaded with a parallel connection of bulk capacitors and inductors.}} \label{RT23}
\end{figure}
\begin{figure*}[t!]
\centerline{\subfigure[{\emph{Material parameters over a wide frequency
range.}}]{\includegraphics[width=7.5cm]{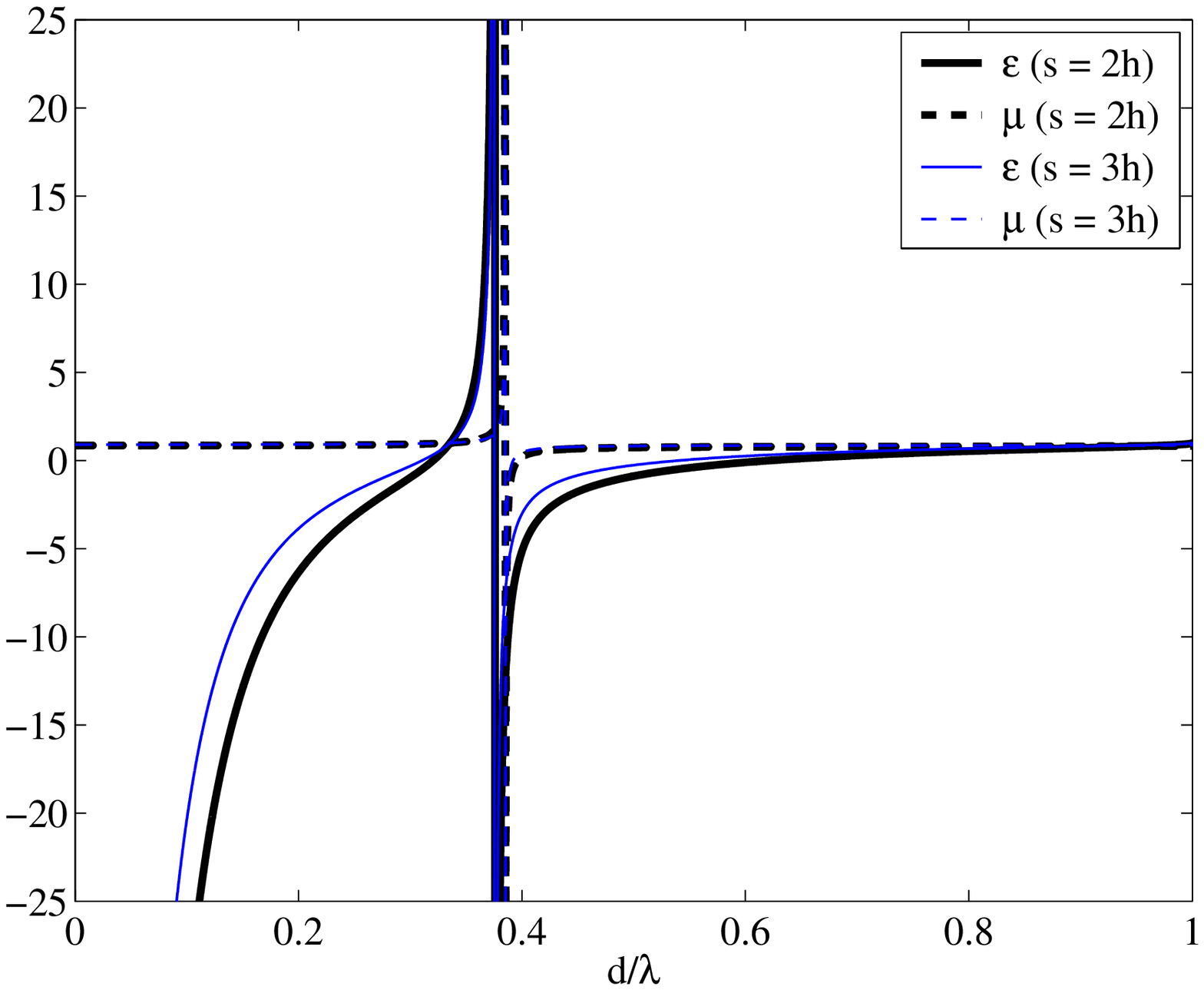} \label{eeff23}} \hfil \subfigure[{\emph{The same result as in
(a), but plotted over a more narrow frequency range.}}]{\includegraphics[width=7.5cm]{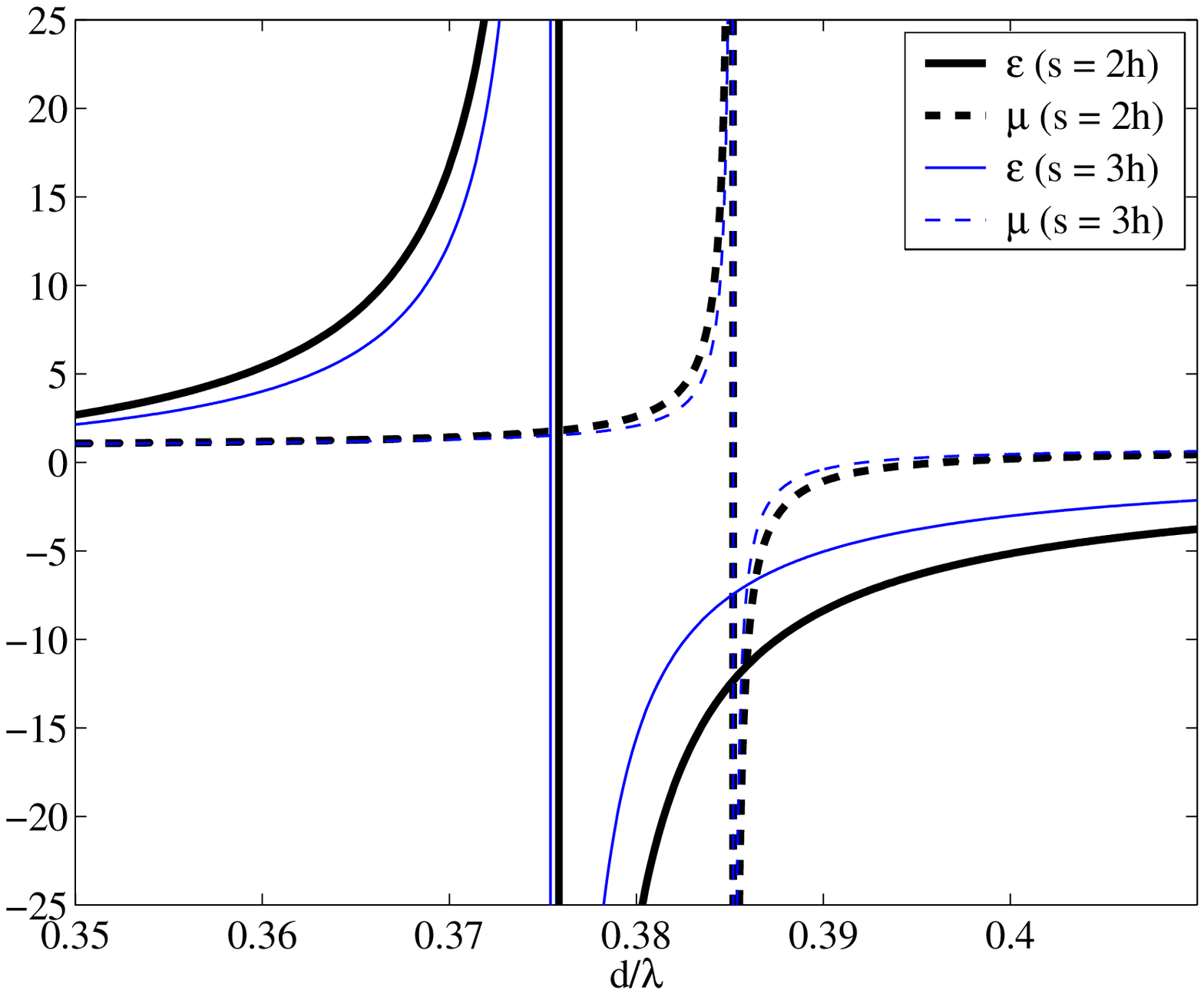}
\label{eeff23_2}}} \caption{{\emph{Mesoscopic effective permittivity and permeability of a double grid of wires
loaded with bulk a parallel connection of bulk capacitors and inductors.}}} \label{eeff23_t}
\end{figure*}

\begin{figure*}[b!]
\centerline{\subfigure[{\emph{Comparing the mesoscopic effective
permittivities.}}]{\includegraphics[width=7.5cm]{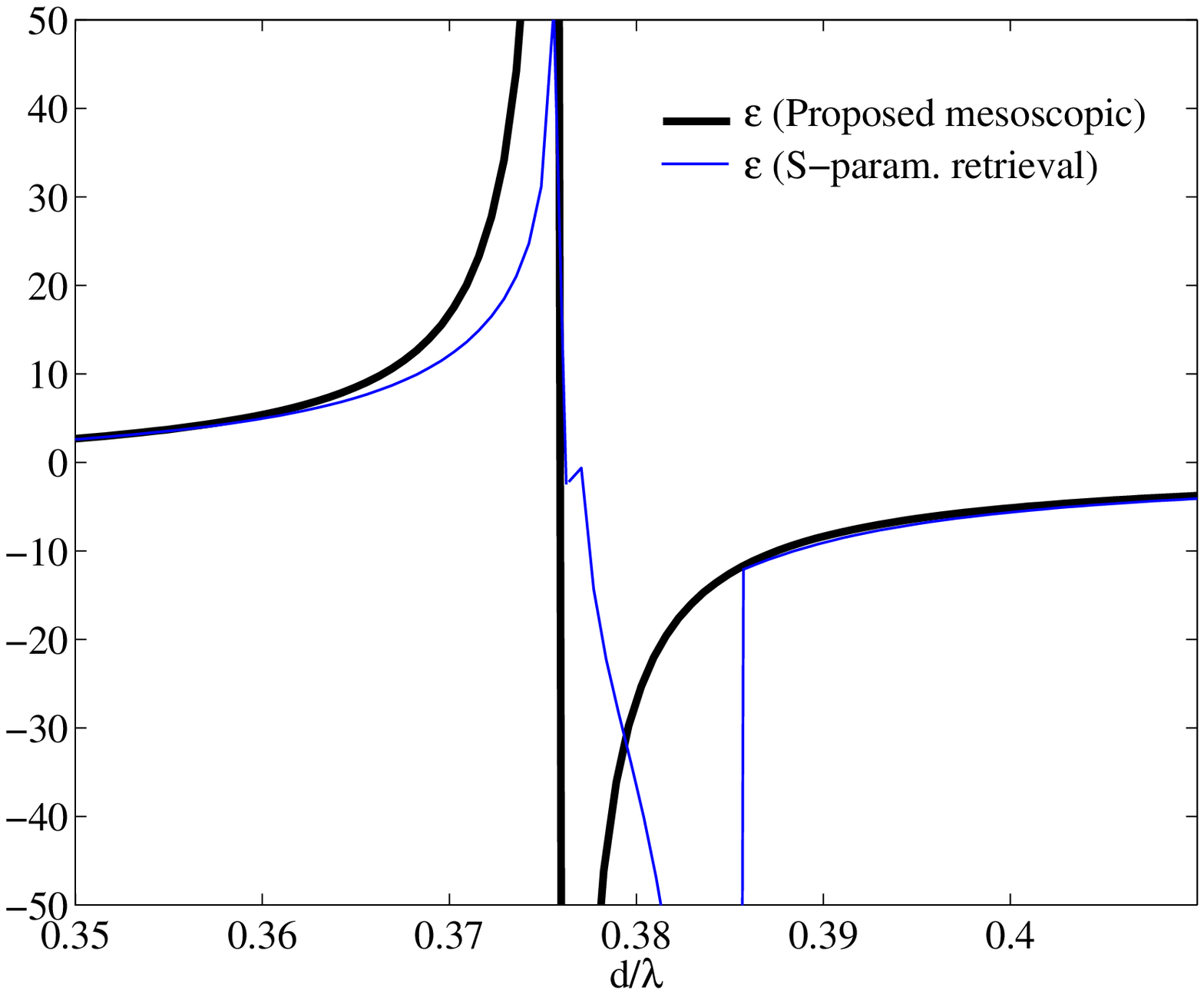} \label{LC_compe}} \hfil
\subfigure[{\emph{Comparing the mesoscopic effective
permeabilities.}}]{\includegraphics[width=7.5cm]{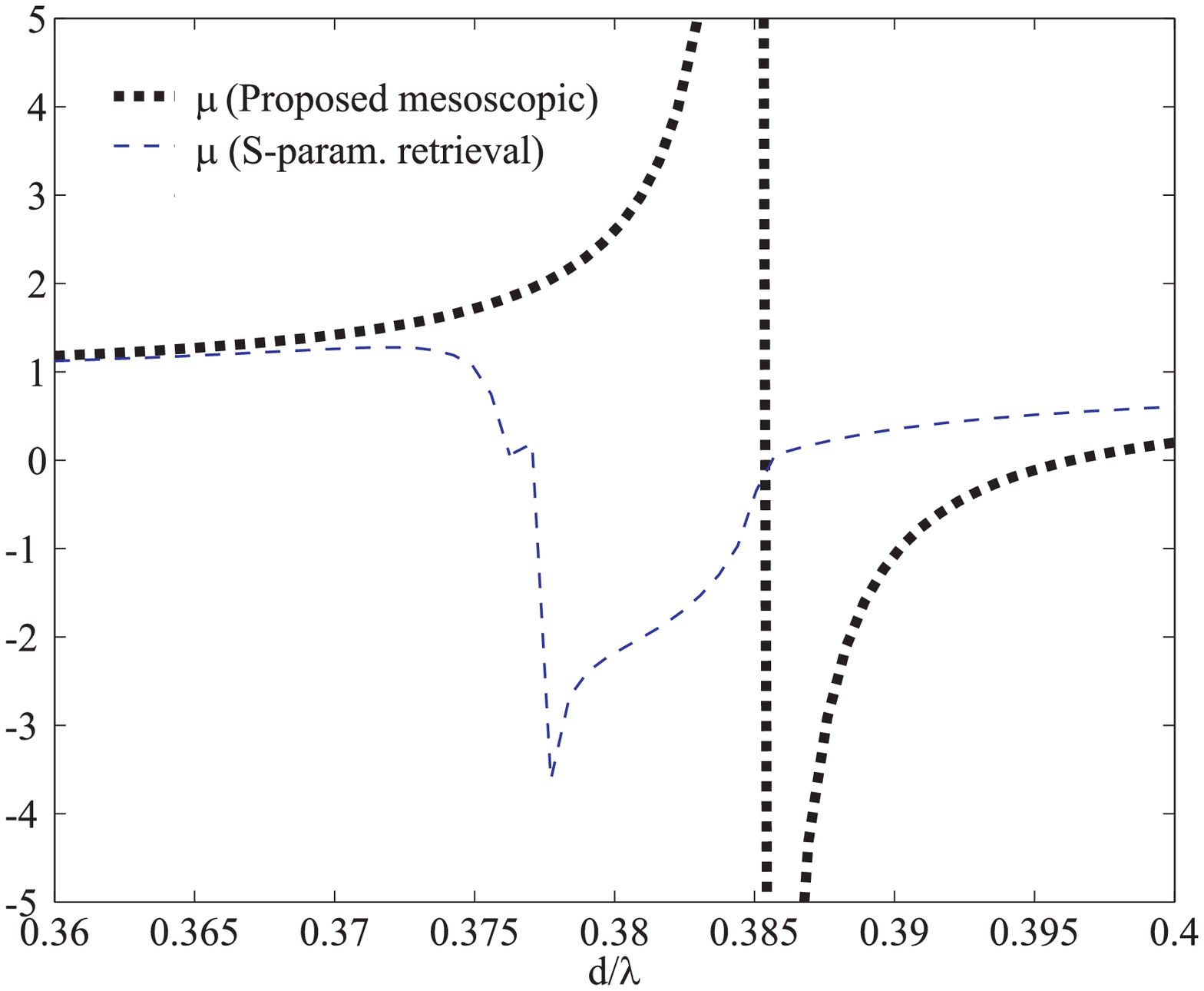} \label{LC_compm}}} \caption{{\emph{Comparison
between the mesoscopic effective material parameters obtained using the proposed method and the $S$-parameter
retrieval method.}}} \label{LC_comp}
\end{figure*}

Finally, we proceed on studying the situation when the wires are loaded with bulk connections of capacitances
and inductances. The reflection and transmission coefficients are shown in Fig.~\ref{RT23}, and they again
resemble the results obtained with the corresponding single grid. However, as explained in the previous
subchapter, there appears a narrow additional transmission window due to interaction between the grids (this
corresponds to the reflection minimum around $d/\lambda\approx0.34$ point).

The mesoscopic effective material parameters are shown in Fig.~\ref{eeff23_t}. All the parameters are purely
real and grow as a function of frequency outside the resonance regions. Qualitative description presented in
connection with the corresponding single grid still holds, however, as is clear from the discussion in the
previous subchapter, the out-of-phase currents flowing in the grids leads to a resonant effective permeability.
Again over a narrow frequency band just above the permeability resonance both material parameters are negative.

\begin{figure*}[t!]
\centerline{\subfigure[{\emph{Unit cell thickness $s=2h$.}}]{\includegraphics[width=7.5cm]{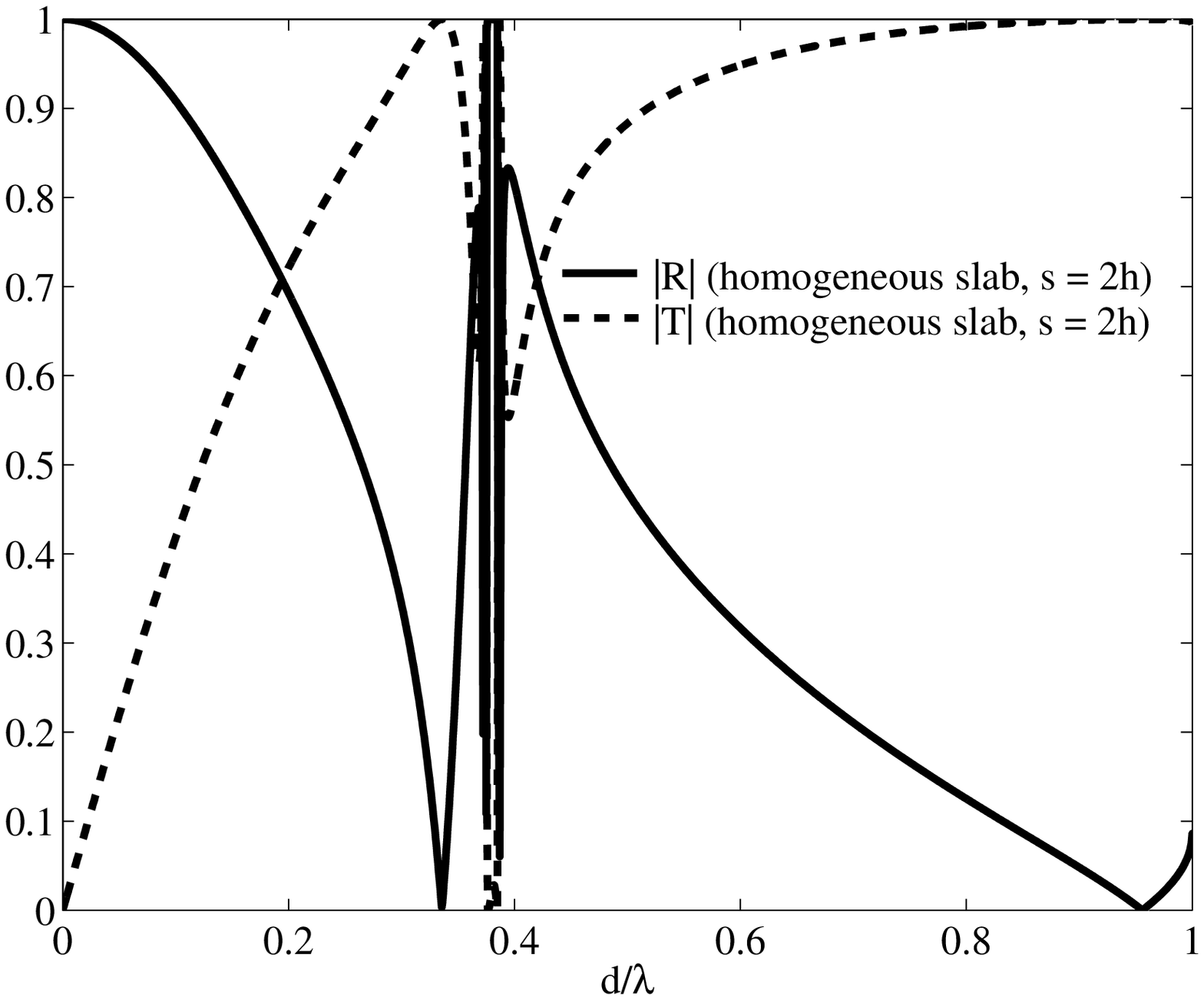}
\label{RT2c31_a}} \hfil \subfigure[{\emph{Unit cell thickness
$s=4h$.}}]{\includegraphics[width=7.5cm]{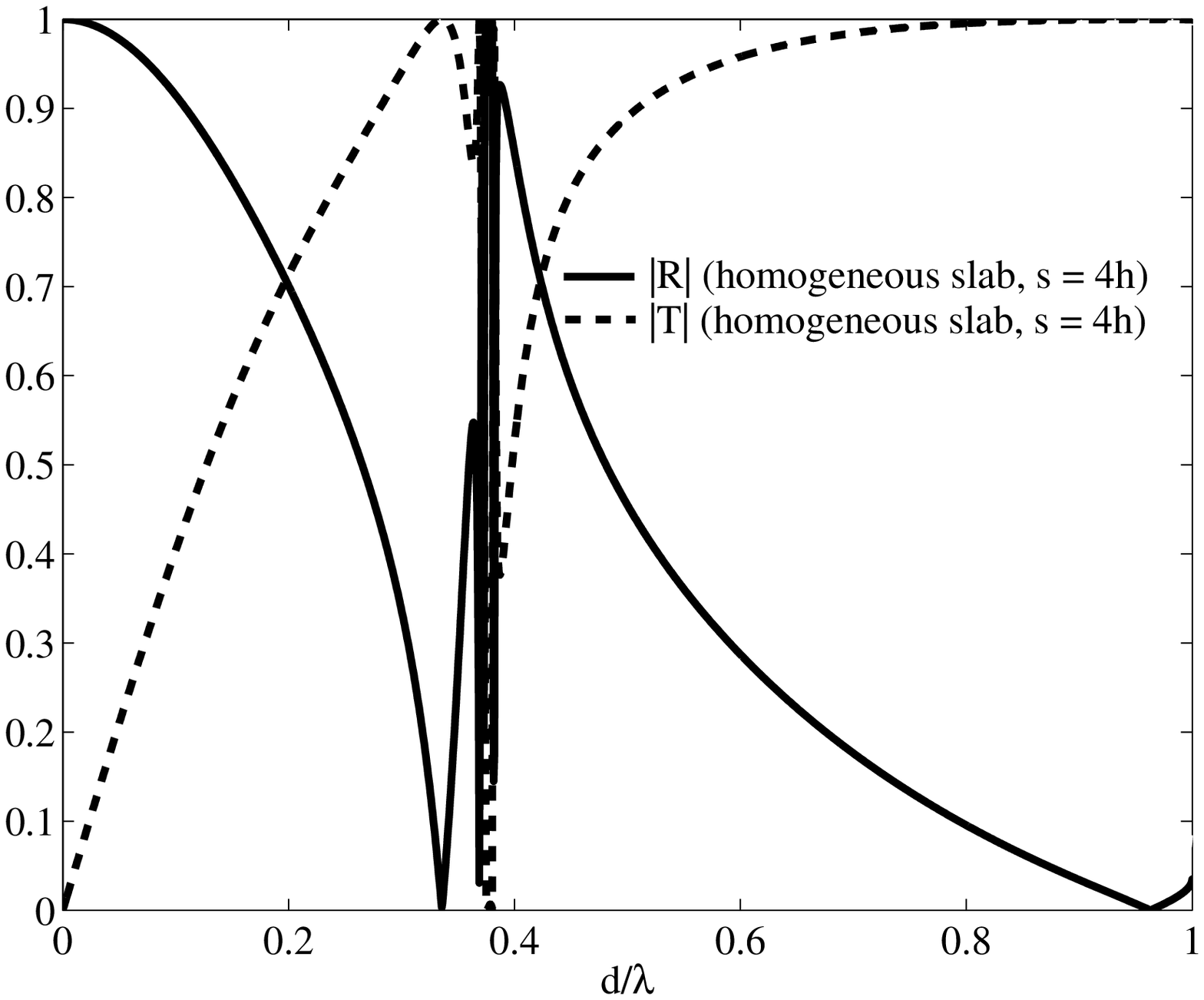} \label{RT2c31_b}}} \caption{{\emph{Comparison between the
reflection data calculated for the actual double grid of $LC$ loaded wires and for the corresponding homogenized
slabs.}}} \label{RT2c31}
\end{figure*}

Fig.~\ref{LC_comp} compares the effective material parameters obtained using the proposed method and the
$S$-parameter retrieval procedure for the case when the slab thickness equals the physical thickness. When
inspecting the parameters obtained through the $S$-parameter retrieval procedure we observe the branch point of
effective refractive index at $d/\lambda\approx0.376$. However, even when excluding the disturbance caused by
the compensation of the branch jump the permeability function, obtained using the $S$-parameter retrieval
procedure, is clearly nonphysical. The reflection/transmission results for the homogeneous slabs (the effective
parameters are those shown in Fig.~\ref{eeff23_t}) are shown in Fig.~\ref{RT2c31}. The behavior is qualitatively
similar to the behavior with a double grid of capacitively loaded wires (please see the discussion in the
previous subchapter).


\section{Conclusion}

We have studied single and double grids of wires loaded with a certain distributed reactive impedance as an
example of thin composite slabs. The currents induced by a normally incident plane wave to all the wires have
been rigorously calculated taking into account the full interaction between the wires. Using the aforementioned
data we have calculated the average polarization and magnetization induced to the structures, and the averaged
fields confined in the unit volume. Using these quantities we have introduced mesoscopic effective material
parameters for the proposed thin composite slabs. A large set of calculated example results has been provided.
The physical background of the introduced mesoscopic material parameters has been discussed in details, and the
proposed results for the double grid structures have been compared with the results predicted by the commonly
adopted $S$-parameter retrieval procedure.

\end{document}